\documentclass[aps,prb,reprint]{revtex4-2}
\usepackage{blindtext}
\usepackage{appendix}
\usepackage{graphicx}
\usepackage{epsfig} 
\usepackage{caption}
\usepackage{soul}
\usepackage{subfloat}
\usepackage[justification=raggedright]{caption}
\usepackage[dvipsnames]{xcolor}
\usepackage[T1]{fontenc}
\usepackage{wrapfig}
\usepackage[english]{babel}
\usepackage{mathrsfs} 
\usepackage{amsmath}
\usepackage{amsfonts}
\usepackage{sidecap}
\usepackage{bbm}
\usepackage[multidot]{grffile}
\usepackage{subfig}
\usepackage{multirow}
\usepackage{url}
\usepackage[english]{babel}
\usepackage{siunitx}

\newcommand{\be}{\begin{equation}}
\newcommand{\ee}{\end{equation}}

\newcommand{\edits}[1]{\textcolor{black}{#1}}

\newcommand{\LTOLFO}{LaTiO$_3$/LaFeO$_3$ }

\newcommand{\LFO}{LaFeO$_3$ }
\newcommand{\LAO}{LaAlO$_3$ }
\newcommand{\YAO}{YAlO$_3$ }
\newcommand{\LuAO}{LuAlO$_3$ }
\newcommand{\LLO}{LaLuO$_3$ }

\begin{document}
\title{From Slater to Mott physics: epitaxial engineering of electronic correlations in oxide interfaces} % in a geological state of iron induced by charge transfer. } %Tuning the strength of electronic correlations by epitaxial engineering : From Slater to Mott physics  in a geological state of iron induced by charge transfer} % This is too long : Maximum 15 words for Nature  
\author{Carla Lupo and Evan Sheridan}

\affiliation{King's College London, Theory and Simulation of Condensed Matter, WC2R 2LS London, UK}
\author{David Dubbink}
\author{Edoardo Fertitta}
\affiliation{Happy Electron Ltd, London W3 7XS, UK}
\author{Chris J. Pickard}. 
\affiliation{University of Cambridge, Department of Materials Science and Metallurgy, Cambridge CB3 0FS, UK}
\affiliation{Advanced Institute for Materials Research, Tohoku University, Sendai, 980-8577, Japan}
\author{Cedric Weber}
\affiliation{ King's College London, Theory and Simulation of Condensed Matter, WC2R 2LS London, UK}

\date{\today}

\begin{abstract} % # words 145
\textbf{Using spin-assisted ab-initio random structure searches, we explore an exhaustive quantum phase diagram of archetypal interfaced Mott insulators, i.e. lanthanum-iron and lanthanum-titanium oxides. In particular, we report that the charge transfer induced by the interfacial electronic reconstruction stabilises a high spin ferrous Fe$^{2+}$ state. We provide a pathway to control the strength of correlation in this electronic state by tuning the epitaxial strain, yielding a manifold of quantum electronic phases, i.e. Mott-Hubbard, charge transfer and Slater insulating states. Furthermore we report that the electronic correlations are closely related to the structural oxygen octahedral rotations, whose control is able to stabilise the low spin state of Fe$^{2+}$ at low pressure previously observed only under the extreme high pressure conditions in the Earth’s lower mantle. Thus we provide avenues for  magnetic switching via THz radiations which have crucial implications for next generation of spintronics technologies.}
\end{abstract}

\maketitle

%$$$$$$$$$$$$$$$$$$$$$$$$$$$$$
%   FIGURE 1
\begin{figure*}[!t]
\centering
\includegraphics[width=0.89\textwidth,height=0.5\textwidth,keepaspectratio]{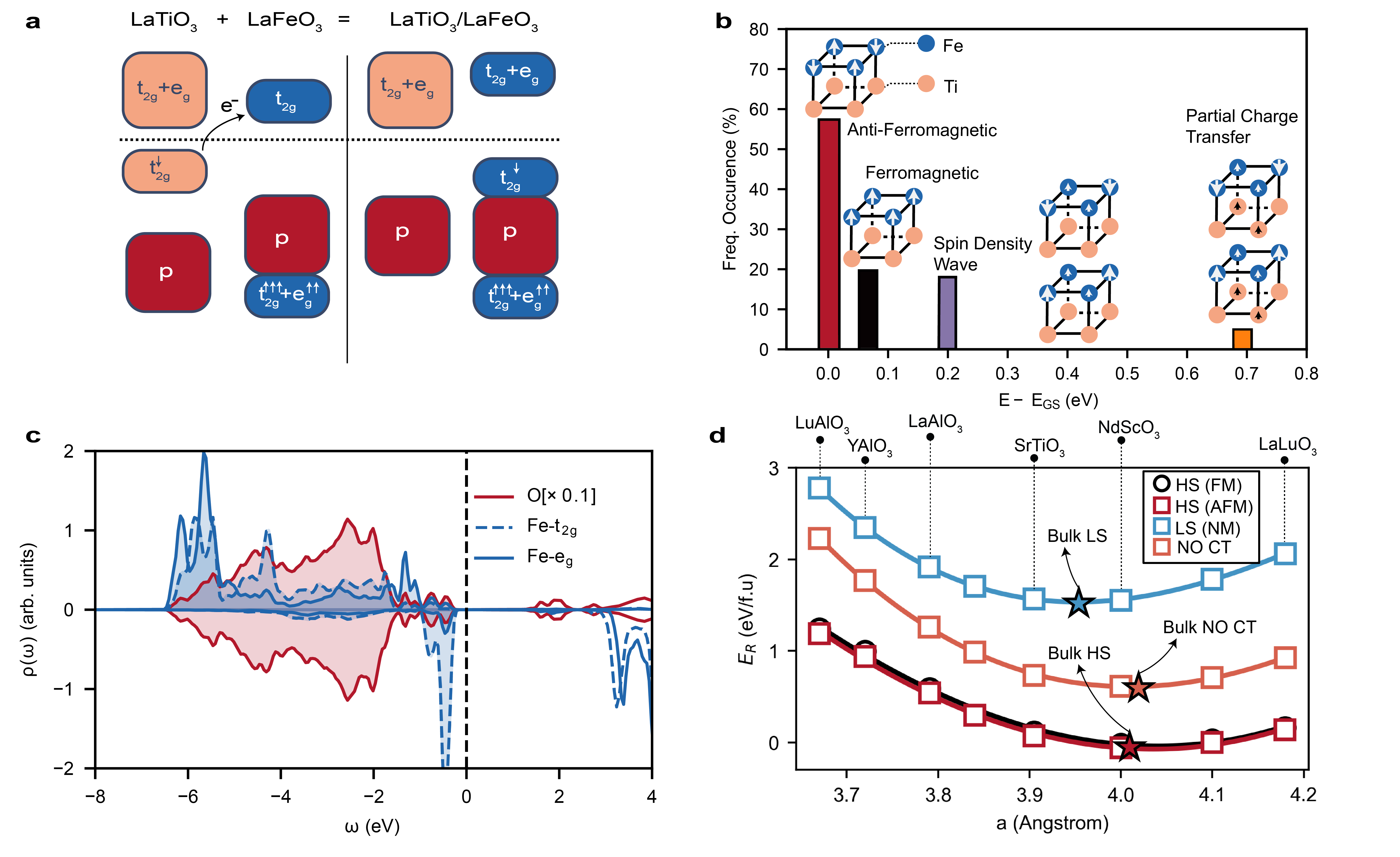}
\caption{\textbf{Epitaxial engineering of robust high-spin ferrous oxides.} \textbf{a} Illustrative band diagrams of constituent materials LaTiO$_3$, LaFeO$_3$ and the superlattice. The dashed black lines are the aligned Fermi levels. The black arrow indicates the direction of transferred charge when forming the superlattice. \textbf{b}  The magnetic phase histogram predicted by spin assisted ab-initio random structure searches of the \LTOLFO superlattice clamped to the \LAO substrate. The reference structure is the fully relaxed bulk \LAO antiferromagnetic \LTOLFO superlattice in the high spin configuration.  The phase space is categorised into 4 different types of magnetic configurations and is ordered by energy.  \textbf{c} Orbital resolved density of states for Fe$_1$ $(3d)$ and O ($(2p)$) majority (positive $y$-axis) and minority (negative  $y$-axis) spin-species at the LaTiO$_3$/LaFeO$_3$ (1/1) superlattice. \textbf{d} % Relative energy ($E_R = E_{\text{sub}} - E_{\text{bulk}}$) stability with respect to the bulk superlattice of the Fe$^{2+}$(S=2) antiferromagnetic ground state compared to the excited Fe$^{2+}$(S=0) nonmagnetic state across different epitaxial oxide substrates. Solid lines are parabolic to the equation of state. 
Relative energy ($E_R = E_{\text{sub}} - E_{\text{bulk}}$) stability  with respect to the bulk superlattice of the Fe$^{2+}$(S=2) antiferromagnetic ground state (red star symbols). Across different epitaxial oxide substrates, a rich set of solutions is provided: charge transfer with Fe$^2+$ in i) low spin (CT LS), ii) high spin antiferromagnetic (CT HS AF), iii) ferromagnetic (CT HS F) configuration \edits{and iv) no-charge transfer (NO CT)}. The star symbols represent the different solutions for the \edits{configuration-dependent ground states} heterostructures in absence of any in-plane strain. }\label{fig:Fig_1}
\end{figure*}
%$$$$$$$$$$$$$$$$$$$$$$$$$$$$$

\section*{Introduction} % WORDS:  should be 500 maximum 

%% I: PI # words 138
Iron compounds and minerals are more often encountered in the ferric (Fe$^{3+}$) than ferrous (Fe$^{2+}$) oxidation state. Under normal conditions the latter is generally observed in its high spin (t$_{2g}^4$ e$_g^2$) state, indeed the distribution of electrons among 3$d$ orbitals for 6-fold coordinated Fe$^{2+}$ results in a high-spin configuration due to Hund's rule and high spin-pairing energy. 
However, established geological models \cite{Li_2004,Speziale_2005,Lin_2012} propose that the extreme pressures deep inside Earth's mantle lead to the collapse of the atomic orbitals of iron from the high-spin to the low-spin state.
In particular, for the case of ferropericlase under increasing pressure a transition towards diamagnetic Fe$^{2+}$ is expected in the pressure range of 40-55 GPa \cite{Badro_2003,Kantor_2006,Lin_2015}. Although the latter transition has been widely studied in minerals, both through measurements and ab-initio calculations \cite{DeGironcoli_2006,Hsu_2011}, a realization of such a transition in synthetic compounds remain unclear.
%% I: PII # WORDS 93
Recent progress in the field of pulsed laser deposition \cite{Ramesh_rev_2019} for oxide nano-engineering  has opened possibilities to study the magnetic transitions of Fe in a controlled fashion. In particular, straining interfaces in oxide heterostructures provides a fertile ground for emergent properties such as ferroelectricity \cite{Vrejoiu_2006,Ramesh_BFO_2004}, high-temperature superconductivity \cite{Schilling_1993},piezoelectricity\cite{Thomas_1997}, magnetoresistance \cite{Tomioka_1995},  structural reconstructions \cite{CJP_I}, multiferroicity \cite{Darrell_2010} and charge transfer \cite{Chen_2017,Zhicheng_2017}. Charge transfer is one important consequence of electronic reconstructions in oxide interfaces \edits{\cite{LTOLFO_PRL,Chen_2013,Weng_2017_ltolvo,Sophie_2019_ltolvo}}. Indeed, in these systems the structural and electronic continuity define a set of band alignment rules, which yield electronic transfer from the high to the low energy bands. 
%% I: PIII # WORDS 103
In correlated systems, where the single electron picture does not hold, charge transfer is closely connected to the strength of electronic correlations. % or Hubbard interactions.
A classification of oxide materials have been provided early on by Zaanen, Sawatzsky and Allen \cite{ZSA_1985}, which distinguishes materials in classes of Mott-Hubbard, charge transfer, and Slater-insulators. The three phases are characterised by different electronic and magnetic states and transitions across them, %, such as %charge localisation-delocalisation transition or spin switching, 
are of great technological interest. Of particular significance in these materials are the low-lying metastable states, manifesting exotic properties attributed to their many body interactions. Exploring this phase space is paramount to fully classifying the practical importance of these materials.
%% I: PIV # words 129
In this study, we report an exhaustive set of exceptional and unprecedented electronic and structural properties in an archetypal oxide superlattice comprising single interfaces of lanthanum-iron and lanthanum-titanium oxides. Using  a spin-assisted ab-initio random structure algorithm \cite{airss_I,airss_II} we predict a multitude of  metastable states and their relative likelihood, and categorise them according to their contrasting magnetic features. We identify a robust anti-ferromagnetic ferrous ground state that can withstand significant amounts of biaxial strain that is influenced by the strength of many body interactions, as a Slater to Mott transition under strain is revealed. Furthermore, we unveil spin state transitions of the ferrous Fe are induced at low pressure by octahedral rotations in the ThZ regime, analogous to what is found at high pressure in the centre of the Earth.

\section*{Results}

%--------------------------------------------------------------
%--------------------------------------------------------------
%                           SECTION 1
%
%                CHARGE TRANSFER AND HIGH SPIN STATE 
%---------------------------------------------------------------

%---------------------------------------------------------------
% Fig1 PI: electronic reconstruction by band diagram  # words 91

In Fig 1a we show a schematic band diagram of the electronic reconstruction at the \LTOLFO interface. 
Considering only the bulk counterparts, we note that Ti and Fe are both 3+ and high spin configuration, with $d^1$ and $d^5$ filling respectively. When the \LTOLFO interface is formed, band alignment of the oxygen states leads to an effective charge transfer of one electron from Ti ($3d$) to Fe ($3d$) states. Thus the charge transfer fixes Ti$^{4+}$ and Fe$^{2+}$ oxidation states in the superlattice, where the ferrous iron has the freedom to occupy either a high- or low-spin configuration. 

%---------------------------------------------------------------
% HERE we should put the Airss discussion done in the SI. # words 194

Using spin-assisted random structure searches we can explore the phase space of the accessible (ground and metastable) magnetic states. We use the AIRSS package \cite{airss_I,airss_II} interfaced with Quantum Espresso to execute the search whose primary constraint is the starting spin configuration. 
% ------discuss the diagram with the statistics 
As shown in Fig 1b, we obtained a classification of the possible spin-states into four categories. The most likely candidates found have an antiferromagnetic ordering, occurring in $57\%$ of cases and are followed by the ferromagnetic structures at 0.07 eV higher in energy, which occur in $20\%$ of instances. The remaining $23\%$ intermediate spin states are characterised by strong spin and charge disproportions, where a manifold of $M_{\pi\pi}$ and $M_{00}$ like \edits{ Spin Density Waves (SDW)} emerge at 0.2 eV. \edits{We emphasise that such SDW metastable configurations, as revealed by our statistical analysis, are not often mentioned in the literature, where the usually studied phases are AFM G-, C-, A- types.} Finally, at 0.7 eV the phase space is dominated by states with only a partial charge transfer occurring, wherein residual moments on one of the Ti ions give rise to a more intricate magnetic ordering. Thus, the spin-assisted random structure search  outlines the ground state of the heterostructure, with charge transfer and Fe$2+$ high spin,  and highlights the importance of the low-lying ferromagnetic configurations, being the most likely metastable state for \LTOLFO with a different magnetic order, \edits{and it ultimately provides a rich phase diagram}(see S.I. for further analysis).

% ------discuss the energies 
%\carla{A more detailed energetic characterization of the different phases reveals that the lowest energies are characterised by antiferromagnetic states of Fe.}
%\carla{The ferromagnetic states lie between 10-100 meV above the ground state. There are some ferromagnetic and antiferromagnetic states that are 100-500 meV above the ground state, and these have unequal magnetic moments on their magnetic ions, and thus exhibit features characteristic of a spin density wave. Finally, the states with spin and charge disproportion are found to lie close to 1 eV above the antiferromagnetic ground state. 
%Thus, the spin-assisted random structure search  outlines the ground state of the heterostructure, with charge transfer and Fe$2+$ high spin,  and highlights the importance of the low-lying ferromagnetic configurations, being the most likely metastable state for \LTOLFO with a different magnetic order (see S.I. for further analysis). }

%---------------------------------------------------------------

% Fig1 PII: description of HS state and energetic argument # Words 210
In Fig 1c we characterise the orbital resolved density of states of the Fe ion. The electron transferred from Ti ($3d$) to Fe ($3d$) sketched in Fig 1a, forms a localized Fe t$_{2g}$ singlet at the top of the valence band. Near the Fermi level (minority, $-0.5$ eV) there is a strongly localized magnetic moment, induced by a spin blocking effect, \edits{where the electronic correlations localise the singlet state by a process where minority spin electron (which ha been transferred from Ti to Fe) can not tunnel to any neighbouring Fe sites, due to their large anti-parallel magnetic moment.} Indeed, the minority Fe $(3d)$ spin is localised by the saturated magnetic moments of its neighbors, a specific feature of the antiferromagnetic ground state in contrast to additional magnetic metastable states (see Fig S1b). 
The optimised high spin configuration can be determined through an energetic competition between the crystal field, super-exchange and Hunds coupling contributions. For an octahedrally coordinated ferrous iron atom coupled to its nearest neighbour in a crystalline lattice, we can consider all density-density interactions $U=V+2J$, crystal field splitting $\Delta$ and nearest neighbour super-exchange coupling. We obtain the energy difference between the low and high spin configuration: $\Delta E^{2+}  = E_{HS}^{2+} - E_{LS}^{2+} = -8J + 2 \Delta - t^{2} / (U + 2J) $. Using typical values of, e.g $\Delta \sim 2 $ eV, $J \sim 0.45$ eV, $U \sim 4.8$ eV and $t \sim 2$ eV, \cite{LTOLFO_PRL} the balance between magnetic interactions and crystal field splitting stabilises the high-spin configuration. Thus, the superlattice inherits the magnetic ground state of the parent \LFO constituent, as the in-plane super-exchange coupling is protected upon formation of the (1/1) interface. \edits{We emphasise that these features persist also in a (2/2) interface ( see Fig.S11 in S.I.)}

%---------------------------------------------------------------
% Fig1 PIII: comparison with previous literature YTO/YFO and La polarization. 

%\carla{MOVE TO SUP MAT (?) A similar result is obtained in previous work where \YTOYFO \cite{ytoyfo} was considered. We note that in the latter work the A-site (Y instead of La) ion has a smaller ionic radius (respectively 104 and 117.2 pm \cite{Shannon} ).  Since the electronic properties are very similar to the materials studied in our work, we conclude that the ionic radius of the A-site ion is not  key to the super-exchange process.  Additionally, a feature that we observed in the La based heterostrucure is an anti-polar distortion of the La toward the FeO$_2$ layer. As shown in Fig 1d, in absence of charge transfer the oxide layers are isopolar (with total nominal charge $\pm1$) and the La is aligned in its oxide plane (LaO). However, when the charge transfer occurs from Ti to Fe, an internal electric field is created, giving rise to an imbalance of charge between the layers. To screen this effect, an antiferroelectric distortion of the La ion of 0.2 \AA   occurs and it prevents further electron transfer by balancing the difference in electrochemical potential between Fe and Ti through aligning their constituent oxygen bands. We confirm that for \YTOYFO the antiferroelectric distortion of the Y ion is enhanced to 0.25 \AA, as expected since the Y ion is lighter than La.}

%---------------------------------------------------------------

% Fig1 PIV:  multisubstrate  analysis and AIRSS  contribution  # words 224
We demonstrate in Fig 1d that through the process of epitaxial engineering the Fe$^{2+}$ high spin configuration is particularly robust to in plane strain. Throughout a range of realistic epitaxial strain amounts (from -5$\%$ to +5 $\%$) there is a consistent $\approx$ 1-3 eV energy difference between magnetic and nonmagnetic configurations. 
Using a spin-assisted random structure search (see S.I. Fig S5 for details) a number of new phases are discovered, the majority being ferromagnetic that lie between 1-10 meV above the antiferromagnetic phase, as shown in Fig 1d.  We notice that, as the in plane states are increasingly localised through tensile strain, then the relative energy difference between the ferromagnetic and antiferromagnetic configurations becomes smaller, which is typical of the onset of a Mott transition. In the opposite limit, where the in plane states are compressed, the states are far in energy, indicative of itinerant magnetism (see Fig S5c). Additionally, two new phases are predicted and illustrated in Fig 1c, the first with No Charge Transfer (NCT) and the other with a Partial Charge Transfer (PCT), both lying about 1 eV above the ground state. We note that the spin assisted random structure search always finds structures with a final configuration that is high spin thus verifying the robustness of the antiferromagnetic ground state, finding that it comprises $\sim$74$\%$ of all predicted structures.

\begin{figure}
\centering
\includegraphics[width=0.5\textwidth,height=0.8\textwidth]{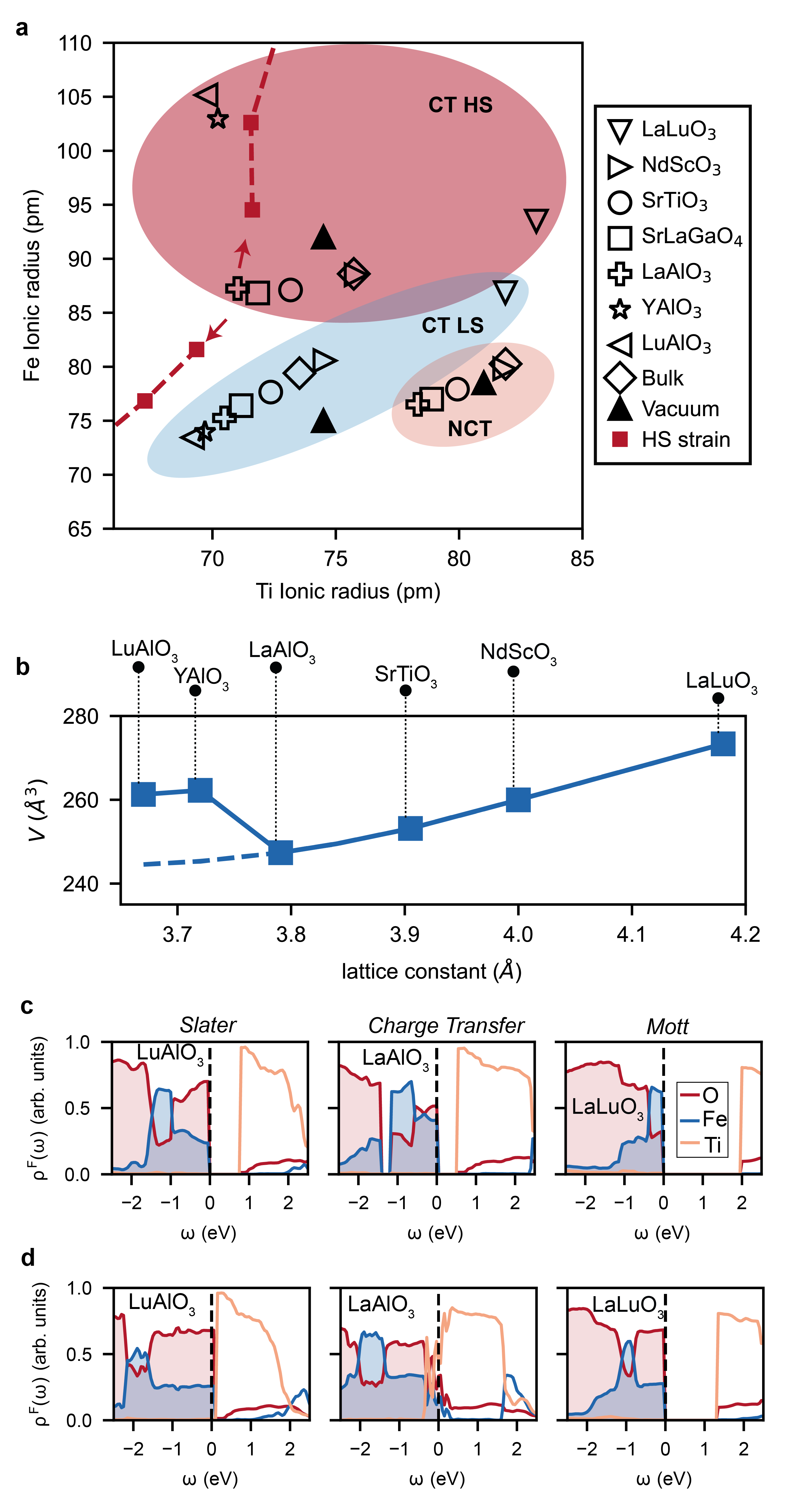}
\caption{\textbf{Ionic radii phase diagram and correlation strength}. \textbf{a} Ionic radii of Ti and Fe for the different configurations highlighted in the shaded regions, respectively for no charge transfer (NCT) with Ti$^{3+}$/Fe$^{3+}_{HS}$ , charge transfer  Ti$^{4+}$/Fe$^{2+}$ where Fe being in i) low spin (CT LS), ii) high spin (CT HS).  The ionic radii obtained of the vacuum configurations (Ref.\cite{Shannon}) corresponding to the nominal valence and spin states of Ti and Fe, are shown as basins of attraction for the different charge/magnetic phases, outlined by the gray shaded regions. The red dashed line connects the configurations obtained under uniaxial compressive strain (along $c$-axis ) starting from the ground state heterostructure with \LAO substrate (star). \textbf{b} shows the dependence of the volume as a function of the selected substrates within the CT HS configurations. \textbf{c (d)}  shows the fractional density of states, for O $2p$, Fe $3d$, and Ti $3d$, relative to the total density for the anti-ferromagnetic HS (ferromagnetic HS) configurations near the Fermi level at 0 for three fixed substrates in the Slater, Charge Transfer and Mott regimes.} \label{fig:ionic_radii}
\end{figure}

%--------------------------------------------------------------
%--------------------------------------------------------------
%                           SECTION 2
%
% IONIC RADII AND THE VOLUME INDUCED SLATER -TO- MOTT TRANSITION
%---------------------------------------------------------------

% -- Fig 2 PI : ionic radii description # WORDS 195
We further extend the analysis of the  phenomena stemming from the epitaxial strain, focusing on the strong coupling between electronic and lattice degrees of freedom. We study as a key structural parameter the ionic radii \footnote{We estimate the atomic radii by averaging over the equatorial and apical bonds of the TiO$_6$ and FeO$_6$ octahedrals of all the compounds shown in Fig (1C)}. Note that we investigate both the ground state and the metastable configurations. Interestingly we notice in Fig (2a) that the obtained ionic radii group together following phase boundaries defined by the charge and magnetic configurations, e.g. Charge Transfer (CT) or No Charge Transfer (NCT), Low Spin (LS) or High Spin (HS). For comparison we include in the diagram the atomic radii obtained in vacuum configurations as in Ref.\cite{Shannon} corresponding to the nominal valence and spin states of Ti and Fe ions (black crosses). The resulting scenario unveils that the latter ionic radii play the role of basins of attraction for the different phases which extend around them. For instance, the values of ionic radii computed for the ground state of the heterostructure (up triangle), are close to the ionic radii obtained in vacuum configurations which are 92 pm and 74.5 pm respectively for Fe$^{2+}_{HS}$ and Ti$^{4+}$ (CT HS)\cite{Shannon}. Similarly all the structures fall close to their nominal atomic configuration.

% -- Fig 2 PII : volume. Expand on the cases of YAO and LuAO # WORDS 214
Outlier cases emerge in the limits of epitaxial compression (\YAO, \LuAO) and expansion (\LLO) in the CT HS configuration, where the ionic radii for Fe has a large effective value for \YAO and \LuAO, whereas the ionic radii of Ti increases for \LLO.  Since the obtained ionic radii  are related to the volumes, in Fig (2b) we analyse its trend as a function of the selected substrates. We notice that a decrease of the lattice constant is coupled to a volume collapse of \LTOLFO with CT HS. However if a $<3.79$ \AA, the ground state volume does not follow the expected trend (dashed line) but instead  expands abruptly, albeit within the stability range of the perovskites (see Fig S7a). A close comparison between the energies of the relaxed configurations and the unrelaxed structure corresponding to the extrapolated volume reveals that volume expansion together with increasing of octahedral rotations (see Fig S7) favour the lowest energy configurations for \LTOLFO constrained on \LuAO and \YAO. In the opposite limit of strain, when the structure is clamped on \LLO the large tensile strain triggers an asymmetric in-plane extension of the Ti-O bonds, which allows the positively charged Ti to effectively screen its neighbouring oxygen anions.  A further analysis of the O-Ti-O angles (Fig S7b), reveals that for substrates with $a>3.905$ \AA, Ti prefers a tetrahedral coordination.

%---  Fig 2 PIII: fractional dos # WORDS 314
We extend the analysis by considering their density of states. The fractional density of states in Fig 2c show that if the heterostructure is clamped on \LLO, the insulating gap is between d(Ti) - d(Fe), hence the configuration is  Mott-Hubbard type. As the in plane lattice constant is reduced to the \LAO one, the band distribution at the Fermi level changes. The Fe $(3d)$ bands, albeit keeping the localised character as in \LLO, are more hybridised with the O $(2p)$, resulting in an overall charge transfer Mott insulator. As the in plane parameter is further reduced (up to \LuAO substrate), the Fe $(3d)$ states show a broader distribution and a larger hybridisation with the oxygen states. A further clarification on the nature of the heterostructure clamped on LuAlO$_3$ is provided looking at the fractional density of states of the ferromagnetic configuration in Fig 2d. We observe metastable metallic behaviour, separated by ~1-10 meV from the anti-ferromagnetic states which are instead insulators. Furthermore we unveil a metal-to-insulator transition depending on the amount of epitaxial strain. We have observed that the superlattice uses a cooperative mechanism between magnetism and structural deformation (volume expansion, increased octahedral rotation and enhanced La  polarization) to lower the energy and stabilise an insulator with itinerant magnetism, bearing similarities with the characteristic features of a Slater insulator. Therefore the CT HS region in the ionic radii phase diagram is characterised by the transition between three different magnetic configurations: from the Slater insulator-like behaviour stemming from the itinerant magnetism to a more localised Mott charge transfer insulator with a O$(2p)$ - Ti$(3d)$ band gap and finally up to a Mott-Hubbard insulating $d-d$ gap. Our results confirm the expected trend that the effect of correlation increases as we go from compressive to tensile epitaxial strain, due to the reduction of the bandwidth. Notably, there is also a simultaneous structural deformation associated to the local coordination of both La and Ti (Fig. S7c). \edits{Moreover, we find that in the Mott phase at room temperature, via paramagnetic Dynamical Mean Field Theory (DMFT) calculations, the charge transfer, local magnetic moment, and spectral weight remains consistent with the the zero temperature DFT magnetic calculations (Fig. S9 and Fig. S10).} 

%--------------------------------------
%--------------------------------------
%           SECTION 3
%
% PRESSURE INDUCED SPIN-STATE CROSS OVER 
%---------------------------------------
 %$$$$$$$$$$$$$$$$$$$$$$$$$$$$$
\begin{figure}
\centering 
\includegraphics[width=0.5\textwidth,height=0.5\textwidth,keepaspectratio]{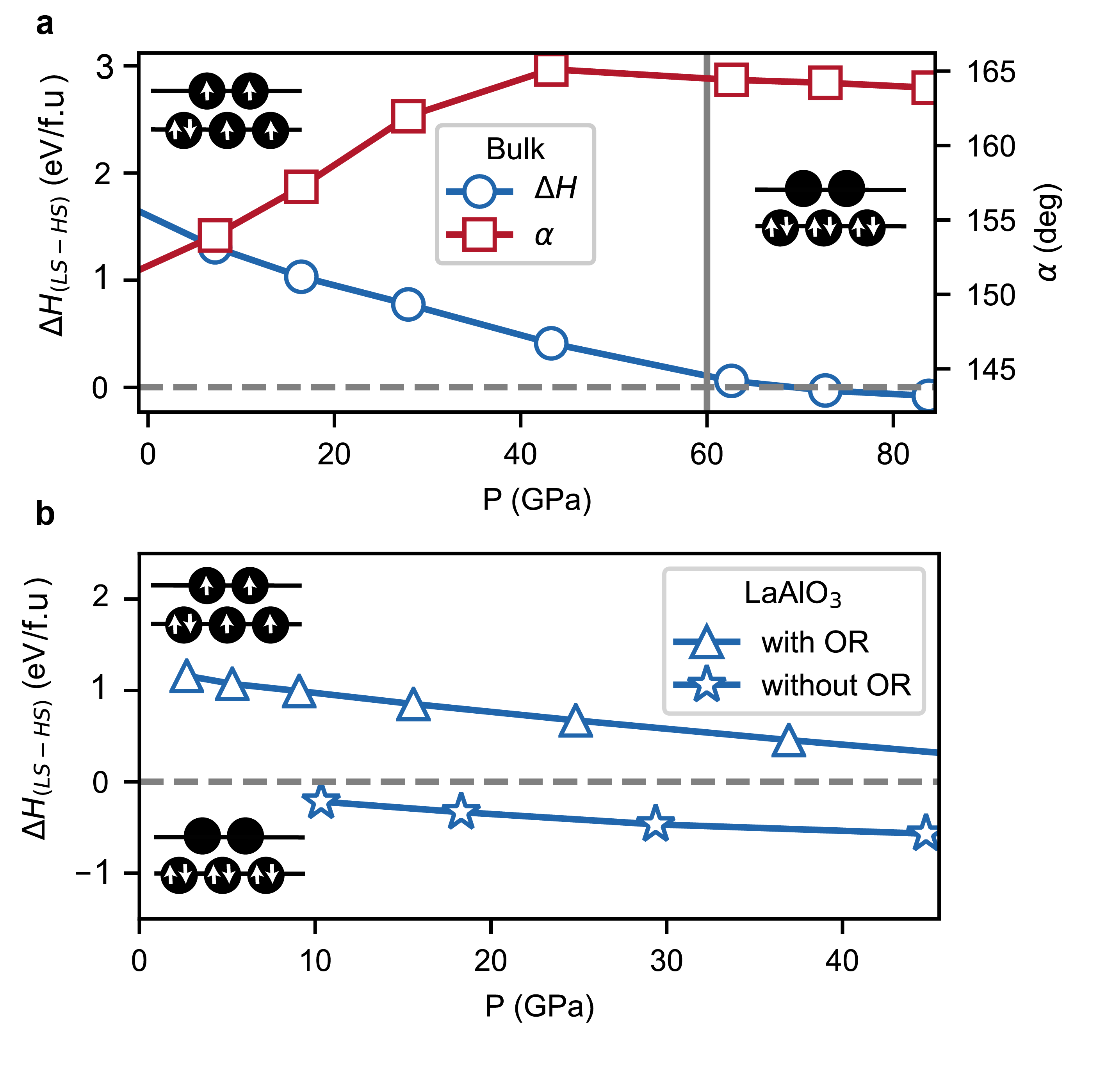}
\caption{\textbf{Pressure induced spin-state crossover} Enthalpy difference between the low spin and high spin configurations obtained under compressive strain along the $c$-axis for \LTOLFO  \textbf{a} bulk and \textbf{b} clamped on the \LAO substrate both with (triangles) and without (stars) octahedral distortions. In panel \textbf{a} the value of the angle $\alpha=$Fe$_1$-O-Fe$_2$ is also shown.}\label{fig:Fig_3}
\end{figure}
%$$$$$$$$$$$$$$$$$$$$$$$$$$$$$

%$$$$$$$$$$$$$$$$$$$$$$$$$$$$$
\begin{figure}
\centering 
\includegraphics[width=0.6\textwidth,height=0.5\textwidth,keepaspectratio]{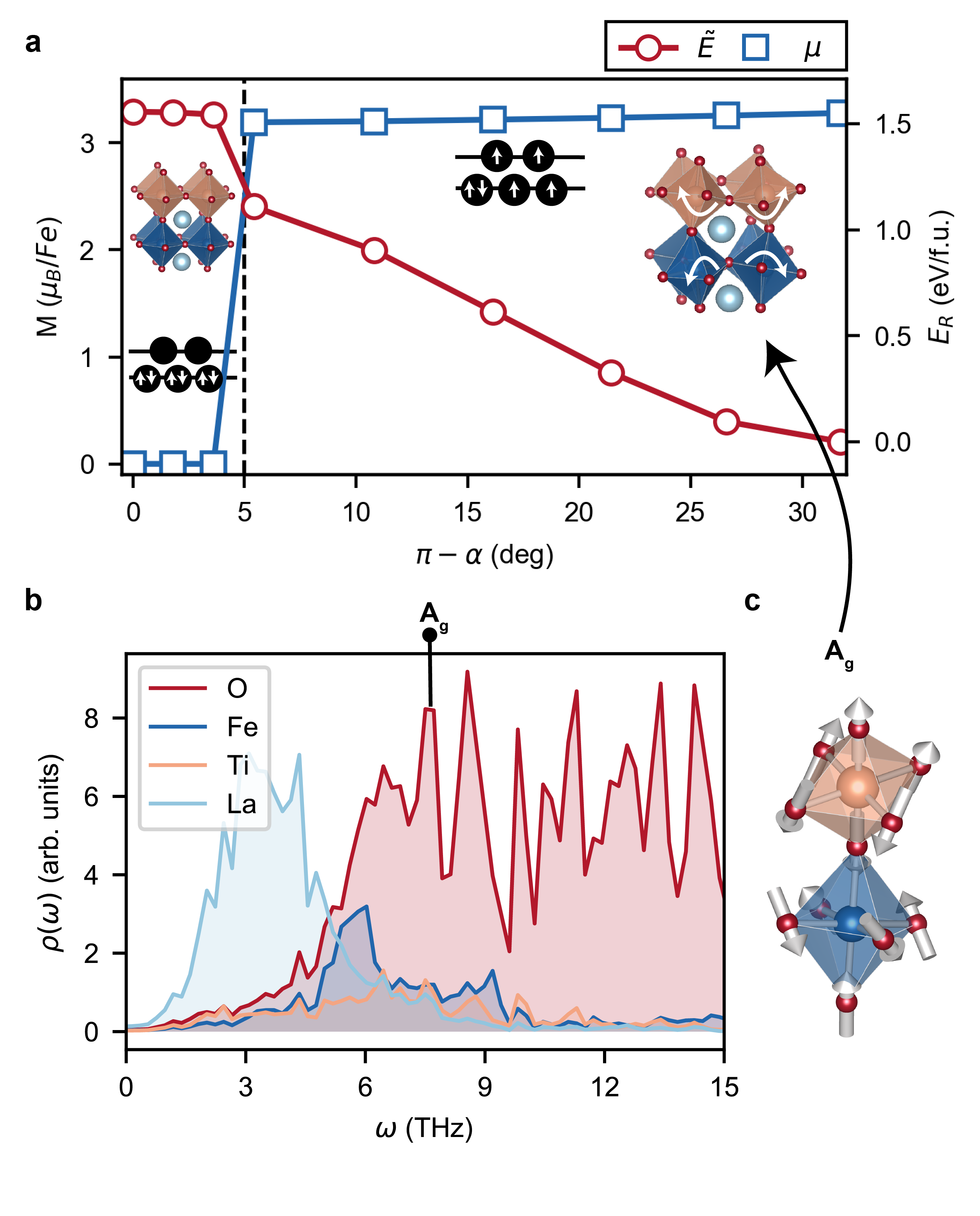}
\caption{\textbf{Terahertz-induced spin state transition mediated by octahedral distortions.} \textbf{a} Magnetic moment per Fe (Relative energy $\Tilde{E}=E_{\text{distortion}}-E_{\rm LaAlO_{3}}$) of distorted structures shown on the left(right) axis, as a function of octahedral rotations $\alpha$. \textbf{b} Atomically resolved phonon density of states for \edits{the structure with $\pi-\alpha=\ang{27}$}. \textbf{c} A$_g$ vibrational mode at 7.42 THz which is compatible with octahedral rotations shown in \textbf{a}.}\label{fig:Fig_4}
\quad

\end{figure}
%$$$$$$$$$$$$$$$$$$$$$$$$$$$$$

% PARAGRAPH 1: BULK UNDER COMPRESSIVE STRAIN ALONG THE C AXIS # Words 320 
We have highlighted that the ground state has a Fe$^{2+}$ high spin configuration, while the low spin one represents a metastable state for all of the structures analysed. An unexplored question that naturally follows, is whether a volume-induced spin crossover can be observed in the heterostructure at hand. The spin crossover in iron has been widely studied in the context of geochemical modelling of the Earth's deep interior.  In particular, crystal field theory and band theory predicted for ferropericlase and iron-bearing magnesium silicate perovskites that a high spin to low spin transition occurs as a result of compression \cite{Li_2004,Speziale_2005,Lin_2012}. Although the latter transition has been widely studied in minerals, a realization of such a transition in deposited compounds remain unclear. Based on similar studies of the structural-induced switching of spin-state ordering \cite{LCO_Rondinelli,LTO_LCO_Twente,LCO_Twente}, we analyse first the stability of the high spin state in absence of substrate under uniaxial compressive strain along the $c$- axis.

For the current analysis there are significant changes in the volume and pressure of the structures studied and energy no longer becomes the thermodynamical quantity of choice, but instead the enthalpy. Thus, in Fig. 3a we show the enthalpies associated with the configurations obtained under uniaxial strain. We predict for increasing compression and consequent volume reduction that the enthalpy difference between the low spin and the high spin configuration decreases, which suggests the progressive destabilization of the latter. In particular, we report that the high spin state remains robust until very large pressure. At a pressure of 60 GPa the system undergoes a transition toward the low spin configuration, as the low spin state accommodates smaller volumes. Furthermore, as a consequence of the uniaxial compression, the octahedral distortions change and we notice that the high to low spin transition is assisted by a quench of the in plane octahedral rotations. This observation reveals that octahedral distortions play a pivotal role in stabilising a low spin configuration in the heterostructure. 

% PARAGRAPH 2: LTOLFO ON LAO UNDER COMPRESSIVE STRAIN ALONG THE C AXIS # Words 131 

We proceed with the comparative study of the enthalpy of structures with and without octahedral rotations shown in Fig. 3b and we consider uniaxial compression (along the $c$- axis) of the \LTOLFO  constrained on \LAO substrate.
Firstly we focus on the low symmetry structure. Despite the introduction of a further in plane pressure, due to the a reduction of $a$ from 4.01 to 3.79 \AA, the configuration with Fe$^{2+}$ high spin remains the most stable one in a pressure range 0-40 GPa. A more interesting picture emerges if we isolate the effect of inducing octahedral rotations. Hence we suppress all octahedral rotations, promoting the starting ground state to a structure in higher symmetry, with non-zero forces only along the z-axis. The configuration with low spin is the most stable structure even at small pressure values.

%-----------------------------------------------
%-----------------------------------------------
%           SECTION 4
%
% OCTAHEDRAL ROTATIONS AND SPIN-STATE CROSS OVER 
%-----------------------------------------------

% Fig 4: # Words 338
Our results so far indicate that the spin transition is weakly affected by the c-axis compression, albeit particularly sensitive to octahedral tilts. We now focus on probing the spin state dependence on octahedral rotations only, in a manner that is independent of strain or pressure and accessible with THz light pumps. Our motivation is linked to advances in state-of-the-art epitaxial growth techniques which have allowed for the atomic control of octahedral rotations, particularly in their capabilities to suppress them \cite{Liao_2016,Liao_2017}. Supplementary to these advances, ultrafast optical spectroscopy can now probe selective low energy phonon modes by a THz pump \cite{Baldini_2019}. We demonstrate that through the manipulation of octahedral rotations in the \LTOLFO superlattice a spin-state transition can be achieved on the \LAO substrate. We analyse the effect of octahedral rotations by constructing a linearly interpolated isovolumetric pathway between the high symmetry undistorted superlattice and low symmetry distorted structures, with in-plane lattice parameters constrained to $a=3.79$ \AA. At each point along the path, the lowest energy spin-state configuration is chosen such that the energy is given relative to the high-spin distorted structure, as shown in Fig 4a. $\alpha$ is the Fe-O-Fe angle and is a measure of octahedral rotation in the superlattice where for $\alpha=\ang{9}$ the spin-state transition occurs. A high energy THz pump can induce lattice vibrations on scales of fractions of an {\AA}ngstrom and this corresponds to the appropriate range defined by $\alpha$. With this in mind, we explore the phonon spectrum, achieved by doing $\Gamma$ point DFPT calculations, to identify vibrational modes within the THz regime that are compatible with octahedral rotations. The phonon density of states, shown in Fig. 4b, is distinctly split between the strongly coupled heavier anions and cations and lighter oxygen ions with a phase boundary at 6 THz. We identify a representative A$_g$ phonon candidate present at 7.42 THz, illustrated in Fig. 4c, that is similar to the tilts induced in Fig. 4a and of primarily oxygen character only. Moreover, there are additional modes beyond 7.42 THz that extend into the Raman spectrum and beyond the range of ThZ spectroscopy.  \edits{In summary, while  the spin-state crossover is hard to achieve through uniaxial compression, since the large ThZ radiation needed would influence structural, dynamic and magnetic stability of the system, a more accessible pathway is provided by manipulating the octahedral rotations which involves transitions at a lower energy scale.}

%\edits {To summarise, while  the spin-state crossovers are hard to achieve when applying uniaxial compression, as the large ThZ radiation needed would influence structural, dynamic and magnetic stability of the system ($\approx 175$ meV), a more accessible pathway is provided by manipulating the octahedral rotations which involves transitions at a much lower energy scale ($\approx 0.05$ eV).}   

%-----------------------------------------------
%-----------------------------------------------
%           CONCLUSION
%-----------------------------------------------
\section*{Discussion} % # words 165
We have shown that a remarkably robust high spin antiferromagnetic ferrous oxide emerges in heterostructure due to interfacial electronic reconstruction. In particular, we illustrate that through epitaxial engineering the electronic correlation strength is tunable and causes a rich phase diagram to arise, with a Mott- to Slater-like transition, \edits{predicted to remain stable at room temperature}. The realization of this effect has direct technology relevance for switchable oxide devices at the nanoscale. Moreover we show that a magnetic phase transition in the superlattice can be triggered by octahedral rotations for which geological scales of pressure are not required. As such, we identify a series of compatible phonon modes at the THz scale in the vibrational spectra that can excite these rotations that are capable of activating the magnetic phase transition. From our results, it is clear that a plethora of fascinating properties emerge beyond what can be attributed solely to electronic interfacial reconstruction alone when additional tunable degrees of freedom are manipulated at the interface of two Mott insulators.\\

\quad

\section*{Methods}
All DFT calculations were performed using the plane wave code Quantum Espresso\cite{QE_2009}, version 6.4.1, together with the GGA-PBE exchange correlation functional\cite{GGA}. Atomic cores were treated within the ultrasoft, nonlinear core correction approach \cite{GARRITY2014446} with valence configurations La(4\emph{f}5\emph{s} 5\emph{p}5\emph{d}6\emph{s}6\emph{p}),  Ti(3\emph{s}3\emph{p}3\emph{d}4\emph{s}), Fe(4\emph{p}4\emph{s}3\emph{d}3\emph{p}3\emph{s}) and O(2\emph{s}2\emph{p}). The plane-wave basis representation is used for the wavefunctions, with a cutoff of 800 eV.  Results are converged by sampling a $6 \times 6 \times 4 $ $\Gamma$-centered k-point mesh in the Brillouin zone. We apply $U=4.8$ eV on Fe and $U=3.0$ eV on Ti within the DFT+U approximation, \edits{where we specify the effective value of the Coulomb interaction  $U_{eff}=U-J$\cite{Dudarev_DFT+U} as implemented in the QUANTUM Espresso package \cite{Cococcioni_2005_LDA+U}}. Our values reflect those commonly used in the literature for the study of a similar system \cite{LTOLFO_PRL}. Both LaTiO$_3$ and LaFeO$_3$ \textcolor{black}{belong to the} \textcolor{black}{\emph{Pnma} spacegroup,}   with  $a^-a^-b^+$ octahedral distortions in Glazer notation. We therefore constructed the (1/1) heterostructure  starting from  unit cell of LaTiO$_{3}$ with 20 atoms which has $\sqrt{2} a_{pc}  \times \sqrt{2} a_{pc} \times 2c_{pc}$ and then  substituting two in-plane Ti ions with two Fe. The heterostructure is then fully relaxed into its (1/1) ground state configuration. We further investigated the stability of the ground state obtained for the unstrained heterostructure, extending the calculation where we apply both uniaxial and epitaxial strain. The former is achieved varying the $c-$ axis of the heterostructure unit cell and allowing for internal relaxation. Regarding the latter, we focus on epitaxial strain constraining the orthorhombic in-plane lattice $a$ and $b$ to the pseudocubic parameter $a_{sub}$ of a set of different substrates, while the out-of-plane lattice parameter $c$ is free to relax. \edits{Therefore, the initial structure for all the simulations with different substrates, characterised by in-plane lattice constant $a_{sub}$, is represented by a 20 atoms unit cell (with 12 oxygens, 2 Fe, 2 Ti and 4 La ions) with the Gd-orthoferrite distortions inherited from the bulk counterparts (LaTiO$_3$ and LaFeO$_3$) and inplane lattice constants constrained to $\sqrt 2a_{sub}\times \sqrt 2a_{sub}$. The allowed relaxations involves internal coordinates of the atoms and the lattice vector along $z-$axis. \footnote{\edits{We clarify that we did not directly include the substrates in the calculations and as a result, we are dealing with only a single interface between two perovskites.}}} Structural degrees of freedom are relaxed until all forces are smaller than 1 mRyd/a.u. In our calculation we consider different initial magnetic conditions for the superlattice in analysis as shown in Fig S2. We calculated the phonon frequencies \edits{ in a ThZ range,} using a finite-difference method \cite{phonopy}. We use the AIRSS package \cite{airss_I,airss_II} interfaced with Quantum Espresso to execute the spin-assisted random structure search. In doing so, we constrain the starting spin configurations and in-plane lattice parameters to a range of epitaxial substrates.\\
\quad
\section*{Data Availability}
The data that support the findings of this study are available from the corresponding authors (carla.lupo@kcl.ac.uk and evan.sheridan@kcl.ac.uk) upon reasonable request.
\quad

% ####################################
% Nature format: up to 50 refs
%####################################

%\bibliographystyle{apsrev4-1}
\bibliographystyle{IEEEtran}
\bibliography{SI.bib}

\section*{Acknowledgements}
CW was supported by grant EP/R02992X/1  from the UK Engineering and Physical Sciences Research Council (EPSRC). C.J.P. acknowledges financial support from the Engineering and Physical Sciences Research Council (Grant EP/P022596/1). This work was performed using resources provided by the ARCHER UK National Supercomputing Service and the Cambridge Service for Data Driven Discovery (CSD3) operated by the University of Cambridge Research Computing Service (www.csd3.cam.ac.uk), provided by Dell EMC and Intel using Tier-2 funding from the Engineering and Physical Sciences Research Council (capital grant EP/P020259/1), and DiRAC funding from the Science and Technology Facilities Council (www.dirac.ac.uk). CL and ES are supported by the EPSRC Centre for Doctoral Training in Cross-Disciplinary Approaches to Non-Equilibrium Systems (CANES, EP/L015854/1). CW is grateful to Antoine Georges for discussions. \\

\quad
\section*{Author Contribution}
 C.L. and E.S. wrote the manuscript. C.L., E.S. and E.F. performed all the calculations. All authors designed the research and revised the manuscript. 
\quad
\section*{Competing interests}
The authors declare no competing interests.\\
\quad
\section*{Additional Information}
Supplementary information has been submitted along with the main manuscript. \\
Correspondence  should be addressed to Carla Lupo (carla.lupo@kcl.ac.uk) and Evan Sheridan (evan.sheridan@kcl.ac.uk).

\end{document}

\title{Supplementary Information\\ From Slater to Mott physics: epitaxial engineering of electronic correlations in oxide interfaces   } 
\author{Carla Lupo*, Evan Sheridan*, Edoardo Fertitta, David Dubbink,  Chris J. Pickard, Cedric Weber}

\maketitle

\section{Magnetic and electronic insights of the unstrained \LTOLFO heterostructure}

%---------------------------------------------------------------
%---------------------------------------------------------------

% Discussion of the different magnetic conditions in the unstrained heterostructure. 
\subsection{Different initial magnetic conditions and outcomes} In our calculations we considered four different initial magnetic conditions (see Fig S1). The G-type I ordering (Fig S1a) represents the configuration with non magnetic Ti ions and antiferromagnetic Fe with in plane Neel ordering. The G-type II ordering, shown Fig S3c, corresponds to the magnetic ordering of both bulk \LTO and \LFO where the interaction between nearest neighbours transition metal is antiferromagnetic both in plane and out of plane.  \\
As shown by the density of states, a different electronic configuration is stabilised according to the initial fixed magnetic pattern.
In particular if a non-magnetic condition is initially set (Fig S1b), in the obtained relaxed \LTOLFO heterostructure we individuate charge transfer from Ti to Fe, with the Fe ions optimised in a 2+ low spin state. \\
If the initial configuration is G-type I, the relaxed electronic/magnetic configuration still shows that the Ti/Fe charge transfer mechanism occurs but with Fe being in a 2+ high spin configuration. In particular the continuous and dashed blue line shown in the density of states, outline that the interaction between the two Fe$_1$ and Fe$_2$ ions is antiferromagnetic. As already stated in the main text, for each Fe ions, the electron transferred from Ti ($3d$) to Fe ($3d$ forms a localized  Fe t$_{2g}$ singlet at the top of the valence band. %A more detailed picture of the energetic arrangement of d orbitals is provided by  Fig \ref{fig:SM_Fig_1a}. We notice that near the Fermi level (minority, $-0.5$ eV) there is a strongly localized magnetic moment, induced by a spin blocking effect. Indeed, the minority Fe $(3d)$ spin is localised by the saturated magnetic moments of its neighbors. The majority  Fe $(3d)$ spins are hybridized with the oxygen at the bottom of the O($2p$) bands at (-6,-5) eV.\\
Interestingly if the initial magnetic order is G-type II (see Fig \ref{fig:SM_Fig_1c}c), the charge transfer is quenched and the band picture of the \LTOLFO heterostructure bears similarities with the bulk \LTO and \LFO counterparts. Indeed we can notice that the $(3d)$ bands of Fe$^{3+}$ (Ti$3+$) are localised below (above) the O $(2p)$ bands, as in the bulk \LFO (\LTO) counterpart. Differently from cases in \ref{fig:SM_Fig_1c} a and \ref{fig:SM_Fig_1c}b, the lower conduction band is mainly characterised by the empty spin channel of the Fe $(3d)$ bands. \\
In the case where an initial ferromagnetic condition (for the Fe ion) is set, with Ti being non magnetic, the density of states obtained in Fig \ref{fig:SM_Fig_1c}d) outlines a solution with charge transfer from Ti to Fe being the Fe1 and Fe2 ions in ferromagnetic order. \\
We clarify that the ground state of the system is represented by the case with charge transfer and Fe$^{2+}$ in high spin configuration \ref{fig:SM_Fig_1c}b, while the other three outcomes represent metastable states at higher energies.

%---------------------------------------------------------------
%---------------------------------------------------------------

\subsection{Orbital and atomic resolved density of states of the ground state configuration} %(Fig S2)}
In Fig. S2 we present the total and orbital resolved density of states for the \LTOLFO superlattice. In particular, we separate the data into three separate panels so as to disentangle the role of the constituent components of the superlattice. In Fig. S2a we see the broad manifold of oxygen states that extend from the Fermi level, where they are primarily hybridised with the d-states of iron, to -7 eV where again there is a significant hybridisation with the states within the iron manifold. Fig. S2b serves to illustrate the empty Ti d-states that locate ~2eV above the Fermi level, and are a superposition of t2g and eg states. Upon forming the interface, the nominal Ti3+ transfers its charge away pushing the Ti-d t2g states 2 eV above the Fermi level, instead of below it. The full electron manifold of the antiferromagnetic Fe states is shown in Fig S2c, where the specific Fe1 and Fe2 states are illustrated in Fig S1a. As mentioned in the main text, we observe spin blocking at the Fermi level where the transferred t2g electron localises at the valence band maximum, with opposite spin orientations for the antiferromagnetic pairs. The sizeable hybridisation with the oxygen states in this region is indicative toward a super-exchange coupling between these magnetic states. Moreover, we remark that the Fe-manifold inherits the main features of its parent compound \LFO, with a d-band splitting above and below the oxygen manifold. To merge the Fe-manifold into itself and destroy the this splitting necessitates the full band to move above the oxygen band, as is observed for the non magnetic scenario in Fig. S1b. In this case, the energy penalty to destroy the AFM state costs ~2eV and hints toward the stabilisation of robust AFM ground state.

%---------------------------------------------------------------
%---------------------------------------------------------------

\subsection{A-site polar distortion: La and Y cases} % (Fig S3)}

We have enphasised so far that the ground state of the heterostructure is characterised by charge transfer from Ti to Fe with Fe in 2+ high spin configuration and with protected antiferromagnetic interactions between the Fe ions in the FeO2 layer. A similar result is obtained in previous work where \YTOYFO \cite{ytoyfo} was considered. We note that in the latter work the A-site (Y instead of La) ion has a smaller ionic radius (respectively 104 and 117.2 pm \cite{Shannon} ).  Since the electronic properties are very similar to the materials studied in our work, we conclude that the ionic radius of the A-site ion is not  key to the super-exchange process.  Additionally, a feature that we observed in the La based heterostrucure is an anti-polar distortion of the La toward the FeO$_2$ layer. As shown in Fig S3, in absence of charge transfer the oxide layers are isopolar (with total nominal charge $\pm1$) and the La is aligned in its oxide plane (LaO). However, when the charge transfer occurs from Ti to Fe, an internal electric field is created, giving rise to an imbalance of charge between the layers. To screen this effect, an antiferroelectric distortion of the La ion of 0.2 \AA   occurs and it prevents further electron transfer by balancing the difference in electrochemical potential between Fe and Ti through aligning their constituent oxygen bands. We confirm that for \YTOYFO the antiferroelectric distortion of the Y ion is enhanced to 0.25 \AA, as expected since the Y ion is lighter than La.

%---------------------------------------------------------------
%---------------------------------------------------------------
\edits{\subsection{Role of $U$}
The crucial mechanism  uncovered in our analysis is the charge transfer between Ti and Fe. In our calculation we set the values of U based on  previous studies of the \LTOLFO interface \cite{LTOLFO_PRL}. In particular $U_{eff}(Fe)=4.8 eV$ and $U_{eff}(Ti)=3 eV$. In Fig.1d of the main manuscript we show a rich phase diagram, where we reported the ground state of the heterostructure and its low-lying metastable states. The small energy differences between the optimised configurations indicate a possible sensitivity of results on different choices of the $U$. In particular, when considering the energy difference between the CT and NO CT case, we should notice that the effective interaction of Fe$^2+$ (d$^6$) HS state is $U-3J$, in contrast to $U+4J$ in the HS state of Fe$^{3+}$ (d$^5$) \cite{Hunds_and_U_Mravlje}. To get further insights in this regard, we extended our analysis to key values of U. In Fig.\ref{fig:U_analysis} we show the  energy dependence of the two configurations  of \LTOLFO interface in interest, while are  charge-transfer high-spin (CT HS, with Fe$^{2+}_{HS}$ and Ti$^{4+}$) and no charge-transfer (NO CT, with Fe$^{3+}_{HS}$ and Ti$^{3+}$) configurations, \st{in} as a function of different values of $U_{eff}(Fe)$. Each data point represents a fully relaxed \LTOLFO heterostructure with fixed $U_{eff}(Ti)=3 eV$ and initial magnetic conditions G-type I and G-type II as defined in Fig.\ref{fig:SM_Fig_1c}. \\
We notice that the CT HS is confirmed to be the ground state of the heterostructure. In particular, for a fixed U, $U_{eff}$ will be smaller for the CT HS and larger for the NO CT. Therefore the energy difference between the two configurations will further increase because of the Hund's J interaction. For example, if we consider $U(Fe)=5 eV$ and $J=0.5 eV$, the effective U for Fe$^{3+}_{HS}$ would be $U_{eff}=U+4J=7 eV$ while for  Fe$^{2+}_{HS}$ we  consider $U_{eff}=U-3J=3.5 eV$ (respectively indicated by the gray symbols). From our results shown in Fig. \ref{fig:U_analysis}, we can estimate that the energy difference is  $\approx 1.5$ eV, therefore the CT HS is the most stable one. }

\section{Epitaxial engineering of many-body correlation strength: multisubstrate analysis}

% Discussion of the Slater to Mott transition via biaxial strain: multisubstrate dos

\subsection{Mott to Slater transition: engineering of itinerant magnetism with compressive strain} % (Fig S4)}
In this section we extend in more details the results showed in Fig 1 and 2 of the main text. 
In Fig \ref{fig:Fig_S4} we report the density of states obtained considering different substrates in their ground state stabilised configuration: charge transfer from Ti to Fe, with Fe$^2+$ in high state state. \\
For each density of state, the two Fe ions in the unit cell are shown with continuous and dashed orange line, highlighting that the long range antiferromagnetic  is protected against potential epitaxially engineered magnetic phase transitions. \\
A closer look at the Fe d states, allows to distinguish different features as increasing in-plane strain is applied. \\
We start considering the substrate with the biggest lattice constant (\edits{\LLO} ). In this case we identify a localised character of the filled Fe-$d$ above the O$-p$. This picture changes as we select a substrate with a smaller lattice constant. In particular with increasing in-plane compressive strain, we can individuate a progressive broadening of the d bands concomitant to an enhanced hybridisation with the oxygen bands, giving rise to an itinerant magnetic behaviour of the Fe ion, typical of Slater insulators. 

% Discussion of the Slater to Mott transition via biaxial strain: fractional density of states 
\subsection{Ferromagnetic solutions: Metal-to-insulator transition } %( Fig S5 )}

Using spin-assisted random structure searching we can explore the phase space around the energy minimum and what intermediate spin-states are accessible. We use the AIRSS package \cite{airss_I,airss_II} interfaced with Quantum Espresso to execute the search. The overall procedure has two primary constraints: (i) the starting spin configuration and (ii) fixed in-plane lattice parameters. As such, the workflow amounts to choosing starting configuration which is then relaxed subject to fixed in-plane parameter from which a ground state configuration is found. In Fig. 1b we show the energetic distribution of total magnetisation per cell clamped to the LAO substrate. What emerges from the search is a classification of the possible spin-states into four categories, in order of likelihood. The most likely candidates found have an antiferromagnetic ordering, occurring in $74\%$ of cases and are followed by the ferromagnetic structures, which occur in $21\%$ of instances. In the remaining cases, which comprise of only $5\%$, are the systems in which there is an overall spin- and charge-disproportion, although there are remnants signatures of an overall magnetic order, depending on the magnitude of the total magnetisation. We note that for these calculations a coarse k-mesh is used, and so the energies should only be trusted qualitatively. We have confirmed that by taking representative candidates from the search that the energy difference between the ground state and predicted antiferromagnetic states is less than 0.001 meV, as expected. The ferromagnetic states lie between 10-100 meV above the ground state. There are some ferromagnetic and antiferromagnetic states that are 100-500 meV above the ground state, and these have unequal magnetic moments on their magnetic ions, and thus exhibit features characteristic of a spin density wave. Finally, the states with spin and charge disproportion are found to lie close to 1 eV above the antiferromagnetic ground state. The spin-assisted random structure search highlights the importance of the low-lying ferromagnetic configurations, being the most likely metastable state for \LTOLFO with a different magnetic order. To this end, we investigate the behaviour of the energy difference between the ferromagnetic and antiferromagnetic configurations for the collection of substrates examined in the main part of the text.

\edits{By extending the spin assisted magnetic structure search beyond \LAO we identify an active polar mode in the titanium octahedral, resulting in a significant change to the local Ti environment. In Fig S6a we present all possible energetically favourable local geometries predicted by the structure search. Upon entering the Mott phase there is a significant Ti-polarisation that lowers the energy of charge disproportioned metastable states, as illustrated with the discovery of a number of different low-lying magneto-structural phases. In the opposite limit on the \LAO substrate the octahedral complex is always stable. Moreover, in Fig  S6b we present the variation of the band gap across the range of substrates studied. The orbital reconstruction of the Ti-d states at the conduction band minimum due to a tetrahedrally coordinated environment shifts the relative energy between the Fe-d states at the valence band minimum. Thus, the increase of the band gap in the Mott phase has a cooperative electro-structural origin. }

\subsection{Structural optimization of the Slater and Mott insulators} % (Fig S6-S7)}
%Discussion of the volume and the dos 
\paragraph{Slater systems under high compressive strain}
We already emphasized in the main text the volume expansion of the Slater systems. In Fig S6 we provide further insights about it and in particular, the E-V plot, shows that the \LuAO and \YAO configurations belong to a different parabola in the landscape of the potential energy surface. 
We also compare the electronic configurations and the energies of the relaxed structure with the ones constrained on an extrapolated volume. 
We notice that the latter are at a higher energy than the former and are characterised by insulating behaviour with reduced band gap  and localised character of Fe $(3d)$ states. 
A closer comparison between density of states of the relaxed configurations and the structure constrained to the extrapolated volume (see Fig S6), reveals at last that volume expansion together with increasing of octahedral rotations (see Fig. S7c), favour the lowest energy configurations for \LTOLFO constrained on \LuAO and \YAO. Moreover, we note that an expansion of the $c$-axis is concomitant with a broadening of the peak of the \eg orbital and increasing hybridisation with the oxygens p bands, giving rise to an itinerant magnetic behaviour of the Fe ion, typical of Slater insulators. 
We note that the volume expansion is only obtained in the magnetic Slater insulator, whereas the non magnetic counterpart is not going through a similar process. Indeed the Slater system provides a mechanism to optimise its kinetic energy, with an associated increase of the La antiferroelectric polarization (see Fig S7b). The latter mechanism is not obtained for Fe$^{2+}$ low spin as its ionic radius is significantly smaller. 

%A further clarification on the Slater nature of the heterostructure clamped on a substrate with small lattice constants (like LuLaO$_3$ and YAlO$_3$), is provided looking at the fractional density of states in Fig \ref{fig:SM_Fig_2b}. In particular we compare the results with stabilised charge transfer and Fe$^{2+}$ in high spin state, being the two Fe ions (Fe1 and Fe2) in antiferro- or ferromagnetic interaction. 
%Interestingly we notice that the ferromagnetic solutions outline a metal-to-insulator transition depending on the compressive strain. Indeed we can observe a metallic behaviour when the selected substrates are \LuAO and \STO, which are instead insulators where antiferromagnetism is stabilised.\\

%\carla{ADD discussion of the energy difference here and connect to the  dos behaviour discussion below}

% Discuss the band gap behaviour
%Thus in order to stabilise a lower energy configuration, the system protect the antiferromagnetic interaction with finite band gap. In particular, looking at Fig \ref{fig:Fig_S4} we note that the band gap shows a decreasing monotonic trend up to LAO and then it suddenly increases for the Slater systems.  

% Discussion of the octahedral rotation and crystal structure stability
%--------------------------------------
% -- Brief discussion Goldschmidt ratio : TO SM
%--------------------------------------
Since the in-plane parameters are fixed to the given substrates, the volume expansion is associated with the increasing of the $c$-lattice parameter which might suggest that the obtained structures are not stable perovskites. 
%To asses the geometric stability and distortions of the crystal structure in analysis, we propose (see Fig. S5b) the analysis of the  tolerance factor (T)\cite{Goldschimdt_1926} in its general definition which can be applied to the heterostructure in analysis \cite{Giustino_2018}. 
To asses the geometric stability and distorsion of the crystal structure in analysis, we propose an extended definition of the Goldscmidth tolerance factor (T) \cite{Goldschimdt_1926}. The $T$ is primarily defined for ABX$_3$ pervoskites, considering the ratios of the constituent ionic radii A, B, X as $T=(R_A+R_X)/\sqrt{2}(R_B+R_X)$. 
In case of \LTOLFO interface, we used a more generalised definition the tolerance factor previously introduced in Ref.\cite{Giustino_2018} and defined for A$_2$BB'O$_6$ double perovskites as $T=2*(R_A+R_X)/\sqrt{2}(R_B+R_{B'}+2*R_X)$. 
Note that the radii for La, Fe and Ti are obtained from the optimised structures while the ionic radius of O$2-$ is fixed to 126 pm \cite{Shannon}.  
In case of an ideal perovskite ABC$_3$ with a cubic closed packed structure the value of the Goldschmidt factor is T=1 (e.g SrTiO$_3$ ). The more T deviates from 1 (t<1), the lower  the symmetry of the stabilised structure is, up to an optimal lower bound for perovskite formability.
On this regard, our result  shows that the volume expansion, which occurs in case of clamping the \LTOLFO on \YAO and \LuAO substrates, is concomitant with a decreasing of the $T$ and hence a further reduction of the symmetry (due to increased octahedral distortions as in Fig. S7c) of the structure, albeit within the stability range of the perovskites.  
Thus we have observed so far that the Slater systems use collaborative mechanism between magnetism and structural deformation (volume expansion, increased octahedral rotation, enhanced La polarization) to lower the energy and stabilises an itinerant magnet.

\paragraph{Mott systems under high tensile strain}
In fig S7(b), we study the behaviour of the angle $\gamma$ defined by Ti and its apical oxygens, and observe that it is constant to $\pi$ up to  \STO. Thus for all the substrates with $a<3.905$, the Ti is in its centrosymmentric position.  When the system is clamped on substrates with lattice parameters greater than 4 \AA, the optimised superlattice is characterized by a non centro-symmetric position of the Ti inside its oxygen cage.

%This plot together with the volume and the Goldschmidt ratio one,  established that LAO substrate lies at the verge of a transition between localized and itinerant magnetism where an associated volume collapse is identified that is concomitant with a Goldschmidt transition and increasing of the band gap.
%We note that the volume expansion is only obtained in the magnetic Slater insulator, whereas the non magnetic counterpart is not going through a similar process. Indeed the Slater system provides a mechanism to optimise its kinetic energy, with an associated increase of the La antiferroelectric polarization (see Fig S5b)

  \section{Temperature dependence of the high spin state}
 \paragraph{Calculation details}
  We perform One Shot (OS) DFT+DMFT calculations at $T=290K=1/40$ eV. The low-energy tight-binding hamiltonian is constructed in the basis of maximally localised Wannier Functions (MLWFs) using the Wannier90 code\cite{w90}. The DMFT self-consistency cycle is implemented using the TRIQS/DFTTools libraries\cite{dft_tools}. For different symmetry inequivalent correlated sites the effective impurity problem is solved using the continuous-time Quantum Monte Carlo hybridisation-expansion solver as implemented in TRIQS/CTHYB\cite{triqs_cthyb}. In this scenario, we can restrict the correlated subspace to that of the Fe$_1$ as a result of the symmetry equivalence at the paramagnetic DFT-level. The double counting correction of Held\cite{HELD_DC} is applied to the correlated subspace as the local interacting Hamiltonian is expressed in the Slater-Kanamori form including spin-flip and pair-hopping processes. The Hubbard U parameter for the Fe$_1$ site is $U=5.3$ eV and the Hund's coupling parameter is set to $J=0.5$ eV, to match the value of the effective Hubbard parameter as used in the magnetic DFT+U calculations.  To reduce noise originating from the Monte Carlo statistics we sample the Legendre coefficients $G_l$ directly within the TRIQS/CTHYB solver. The full frequency spectral function $A(\omega)$ is obtained via analytically continuing the Matsubara Green's function using The TRIQS/MaxEnt library\cite{triqs_maxent}. All results are performed within the "one-shot" DFT+DMFT approximation and neglect the effect of charge self-consistency. The starting configuration used to generate the low energy effective Hamiltonian is a paramagnetic DFT calculation without Hubbard U correction where the seed structure is a converged LSDA+U calculation on the LaLuO$_3$ substrate.
  
  \paragraph{The effect of temperature}
  In Fig. S9(a) we illustrate that the magnetic Fe$^{2+}$ state is recovered at room temperature within the paramagnetic DMFT approximation and predicts the same orbital ordering as the magnetic DFT prediction. We also emphasise in Fig. S9(a) that paramagnetic DFT is only capable of obtaining a low-spin non-magnetic configuration, with fully occupied $t_{2g}$ levels. Moreover, in Fig. S9(b) we show the converged imaginary part of local interacting orbital resolved matsubara Green's function, $\Im m G(i \omega_n)$,  which attains an insulating-like character at zero frequency. Similarly, the imaginary component of the converged impurity orbital resolved self energy $\Im m \Sigma(i \omega_n)$ is presented in Fig. S9(c) and shows a clear distinction between the build-up of interactions within the strongly correlated Fe-d impurity states. Notably, the fully-occupied $d_{yz}$ has as significantly flatter self-energy than its counterparts, as should be expected in an almost full orbital\cite{medici_caponeFe2+}, while the additional orbitals all attain a negligble quasiparticle weight, as an indication of their localised character. The retention of the orbital reordering of the Fe$^{2+}$ state at $T=290K$ within the DMFT approximation demonstrates that many-body correlations persist and can mediate long-range magnetic order beyond the zero temperature approximation of magnetic DFT+U. \\
  
% Magnetic moment   
  To assess the dependence of magnetic moment of Fe at room temperature we directly sample the fluctuating paramagnetic moment $S = \sqrt{\langle S_{z}^{2}\rangle}$ where $S_{z} = \frac{1}{2} \sum_{\mu} (n_{\uparrow}^{\mu} - n_{\downarrow}^{\mu}) $ is the total z-component of the isotropic magnetic moment summed over the Fe-d orbitals. From this, we estimate that the local magnetic moment at room temperature is approximately $m_{Fe}^{DMFT} = 2.82 \mu_{B}$. For zero temperature magnetic DFT we estimate that the magnetic moment of the high spin Fe to be $m_{Fe}^{DFT} = 3.46 \mu_{B}$. Thus, we see that there is an overall reduction of the magnetic moment by $~18 \%$ at room temperature due to thermal and quantum fluctutations. Since the DMFT calculation treats the  the fully rotational invariant vertex with the non-diagonal terms of the Hund’s coupling (whereas in DFT only the diagonal terms are included), we also anticipate there is a reduction of the magnetic moment due to additional scattering terms, neglected in the vanilla DFT. \\
  
% Spectral weight
  Through analytical continuation of the converged $G(\tau)$ using the maximum entropy technique the spectral function is obtained at $T=290K$, as illustrated in Fig. S10 decomposed along the crystal field axes. We also provide a comparison to the $T=0K$ scenario obtained with magnetic DFT+U clamped to the LaLuO$_3$ substrate. It is also observed in the spectral function that the Fe$^{2+}$ paramagnetic state remains stable at higher temperatures and sustains the orbital ordering predicted by zero temperature magnetic DFT+U. However, we note that there is significantly less splitting between the upper and lower Fe-d valence bands within the DMFT approximation. As already discussed in relation to the magnetic moment, we attribute this to the dynamical spin effects captured by the Slater-Kanamori Hamiltonian. In addition, we remark that the maximally localised wannier functions are obtained only in the window of energy of the Fe-d manifold subsapce, and we neglect the contributions arising from the O-p states. As such, this will have an additional effect on suppressing the overall splitting of the Fe-d states\cite{BeckLVO,Chen_2017}, due to the absence of O-p alignment across the interface in the Wannier basis. Moreover, we see that the Fe-d gap persists up to room temperature, indicative of room temperature Mott stabilisation of the magnetic insulator.

\section{(\LTO)$_2$/(\LFO)$_2$ heterostructure}
We extended our calculation to larger supercells. In particular we considered the  (LaTiO3)$_2$/(LaFeO3)$_2$ heterostructure as shown in Fig.\ref{fig:22_supercell}. Following the same procedures as for the (1/1) heterostructure, we optimised different configurations starting from a non-magnetic and a G-type II initial magnetic ordering. The density of states shows that, irrespective of the initial conditions, a charge transfer from Ti to Fe occurs, stabilizing Ti$^{4+}$ and Fe$^{2+}$. Indeed the Ti d states characterise the conduction bands in the proximity of the Fermi level. On the other hand, the Fe-d states characterise the top of the valence bands, assuming a different character according to the stabilised spin state of the Fe$^{2+}$. Indeed in panel b) we identify a low spin state of both Fe1 ad Fe2, with filled t$_2g$ states, while in panel c) the lifted symmetry of Fe1 and Fe2 indicate their antiferromagnetic interaction and for each Fe ions, we retrieve the characterization of the d manifold already observed in Fig 1c of the main text. We emphasise that also in the case of the 2/2 interface, the ground state is CT HS and the non-magnetic configuration is a metastable state higher in energy by $\approx 1.6$ eV/f.u.
Therefore, most of the observed features bear similarity with the (1/1) interface, although in this case, we can observe a reduced bandgap. 
Interestingly, a comparable result has been observed for the \YTOYFO interface in Ref. \cite{ytoyfo} (Fig 2 in Ref.\cite{ytoyfo}), where the authors find that even if increasing the number of layers, charge transfer from Ti to Fe occurs, resulting in a high spin state for the Fe2+. Similarly to our case, the electronic properties of the 1/1 and 2/2 interface differ for the width of the d-d bandgap, which is reduced in the latter.
Lastly, it is worth highlighting the structural deformation stemming from the cooperative charge-lattice degrees of freedom. Indeed, irrespective of the spin-state optimised, we observed the La anti-polar distortion triggered by the charge transfer from Ti layers to Fe layers. This effect becomes even more interesting for this larger heterostructure. We can differentiate both aligned and misaligned LaO layers. In particular, La is aligned with its oxygen plane if in between a symmetric potential, as observed for the LaO layers in between the double \LTO or \LTO. Instead, if the LaO is at the centre of asymmetric potential, the La ions move toward the more negatively charged layer, which in this case is represented by FeO$_2$.
\begin{figure}[t]
\centering
\includegraphics[width=\textwidth,height=1.2\textwidth,keepaspectratio]{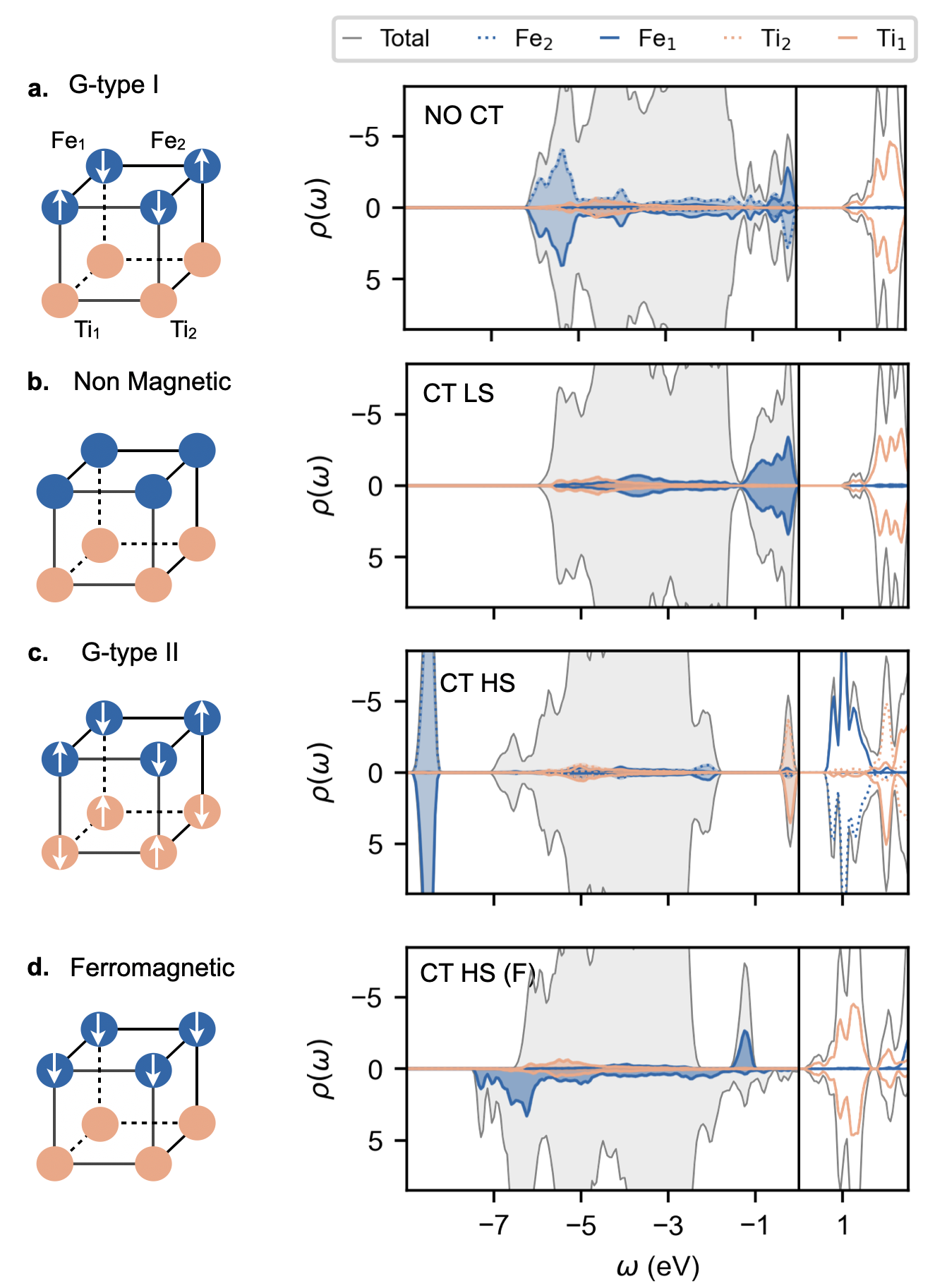}
\caption{The total and species resolved density of states for (a) G-type I (b) Non-Magnetic, (c) G-type II antiferromagnetic \edits{and (d) ferromagnetic (only Fe ions)} configurations  of the (LaTiO$_3$)$_1$/(LaFeO$_3$)$_1$ superlattice. In every instance, the full Ti-d and Fe-d manifolds are shown and  are decomposed into their Fe$_{1,2}$ or Ti$_{1,2}$ spatial labels. }\label{fig:SM_Fig_1c}
\end{figure}

%%%%%%%%%%%%%%%%%%%%%%%%%%%%%%%%%%%%%%%%%%%%%%%%%
% FIG: unstrained LTOLFO, CT HS, details orbitals
%%%%%%%%%%%%%%%%%%%%%%%%%%%%%%%%%%%%%%%%%%%%%%%%%
\begin{figure}
\centering
\includegraphics[width=\textwidth,keepaspectratio]{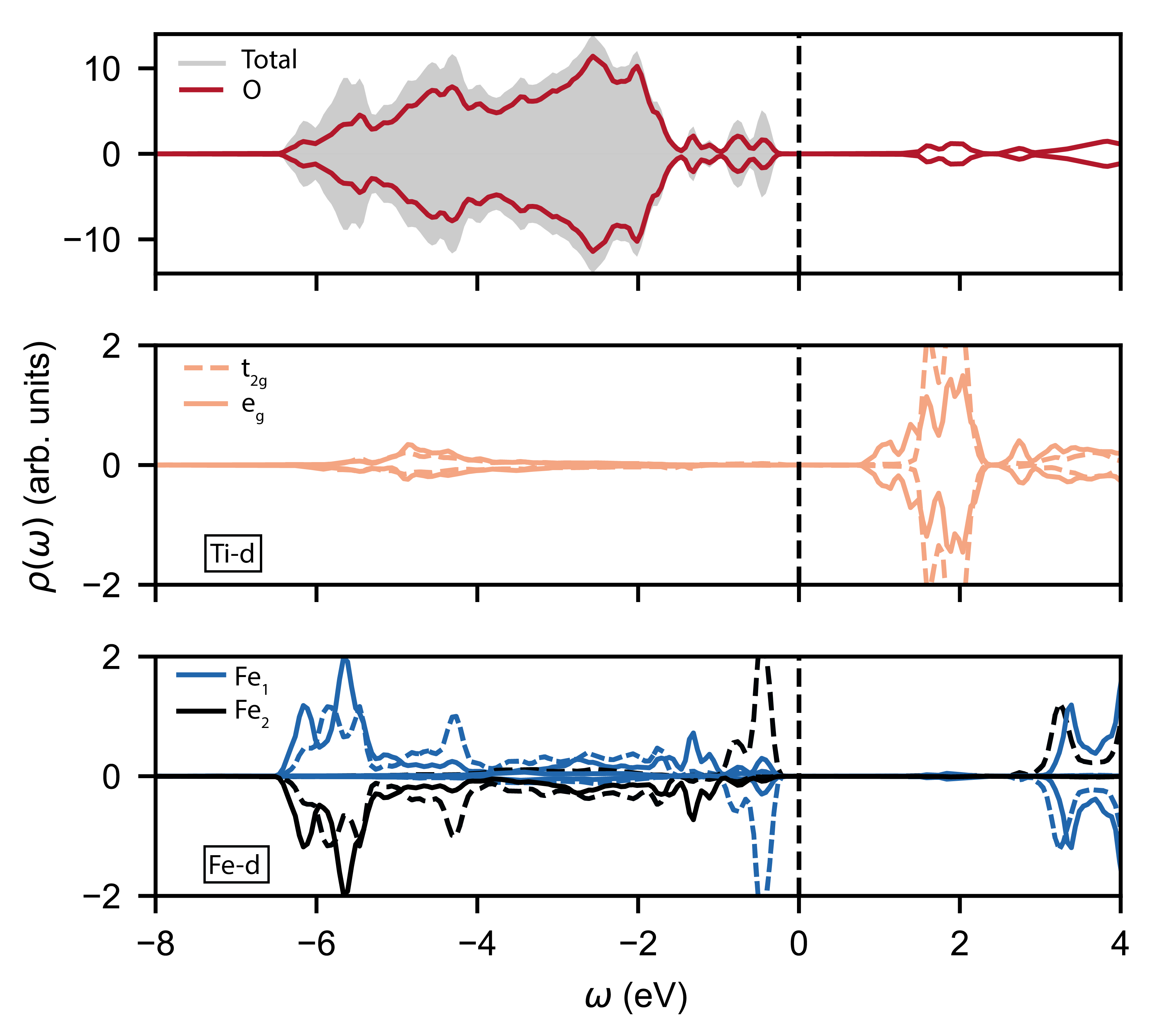}
\caption{(a) The total and orbital resolved density of states for the O p majority (above axis) and minority (below axis) spin-species at the (LaTiO$_3$)$_1$/(LaFeO$_3$)$_1$ superlattice. (b) The orbital resolved density of states for the full Ti electronic manifold, decomposed into $t_{2g}$ and $e_g$ components. (c) The orbital resolved density of states for the full Fe electronic manifold, decomposed into $t_{2g}$ and $e_g$ components and distinguished by atomic label.}\label{fig:SM_Fig_1a}
\end{figure}

%%%%%%%%%%%%%%%%%%%%%%%%%%%%%%%%%%%%%%%%%%%%%%%
% FIG: unstrained LTOLFO, La  polar distorsion
%%%%%%%%%%%%%%%%%%%%%%%%%%%%%%%%%%%%%%%%%%%%%%%
\begin{figure}[t]
\centering
\includegraphics[width=\textwidth,keepaspectratio]{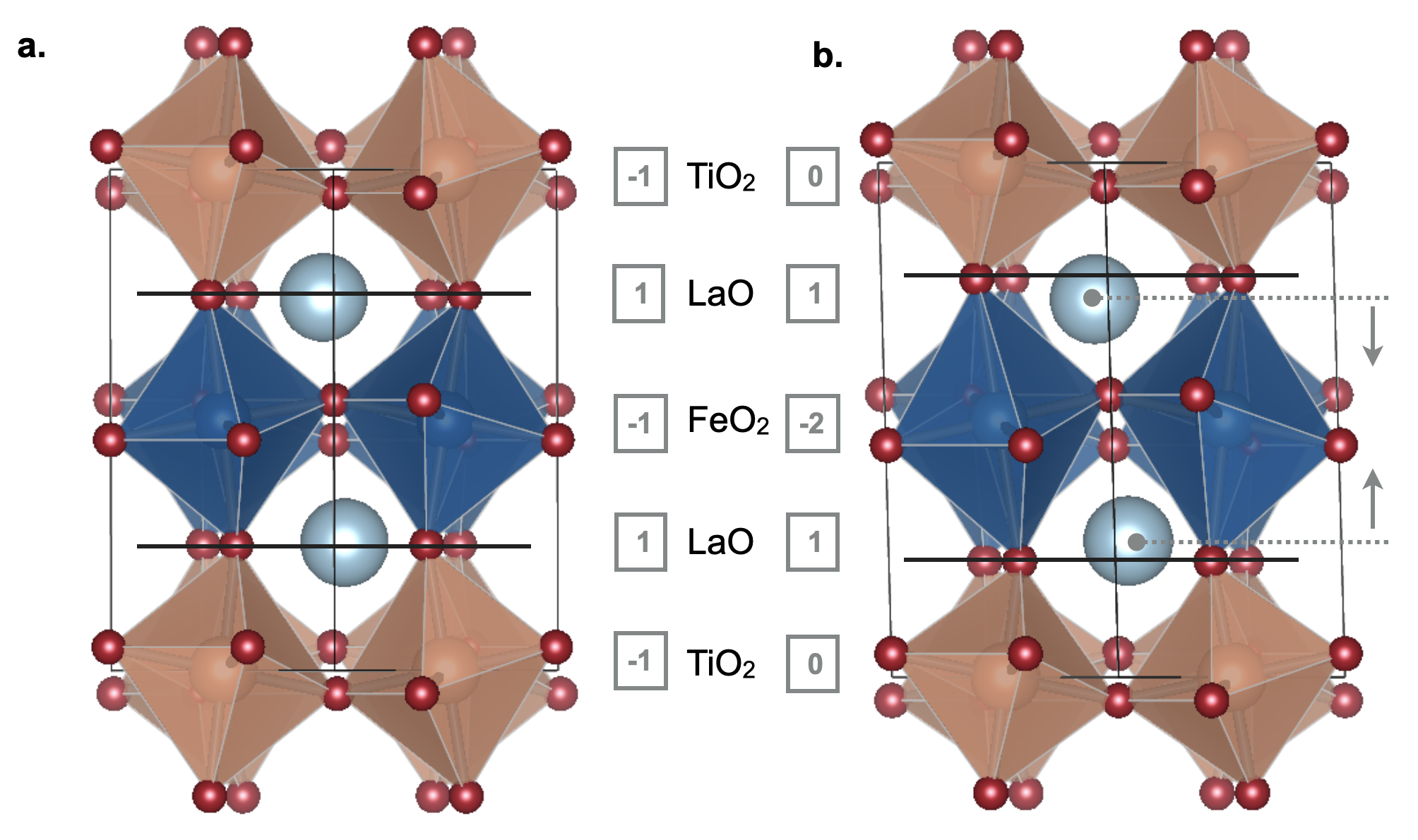}
\caption{The LaTiO3/LaFeO3 heterostructure with electronic reconstuction at the interface (a) without charge transfer and (b) with charge transfer. For each case, we
identify the nominal charge of the associated layers and the La-O displacement.}\label{fig:SM_Fig_1c}
\end{figure}

%%%%%%%%%%%%%%%%%%%%%%%%%%%%%%%%%%%%%%%%%%%%%%%%%
% FIG: U_Eff analysis 
%%%%%%%%%%%%%%%%%%%%%%%%%%%%%%%%%%%%%%%%%%%%%%%%%

\begin{figure}
\begin{center}
\includegraphics{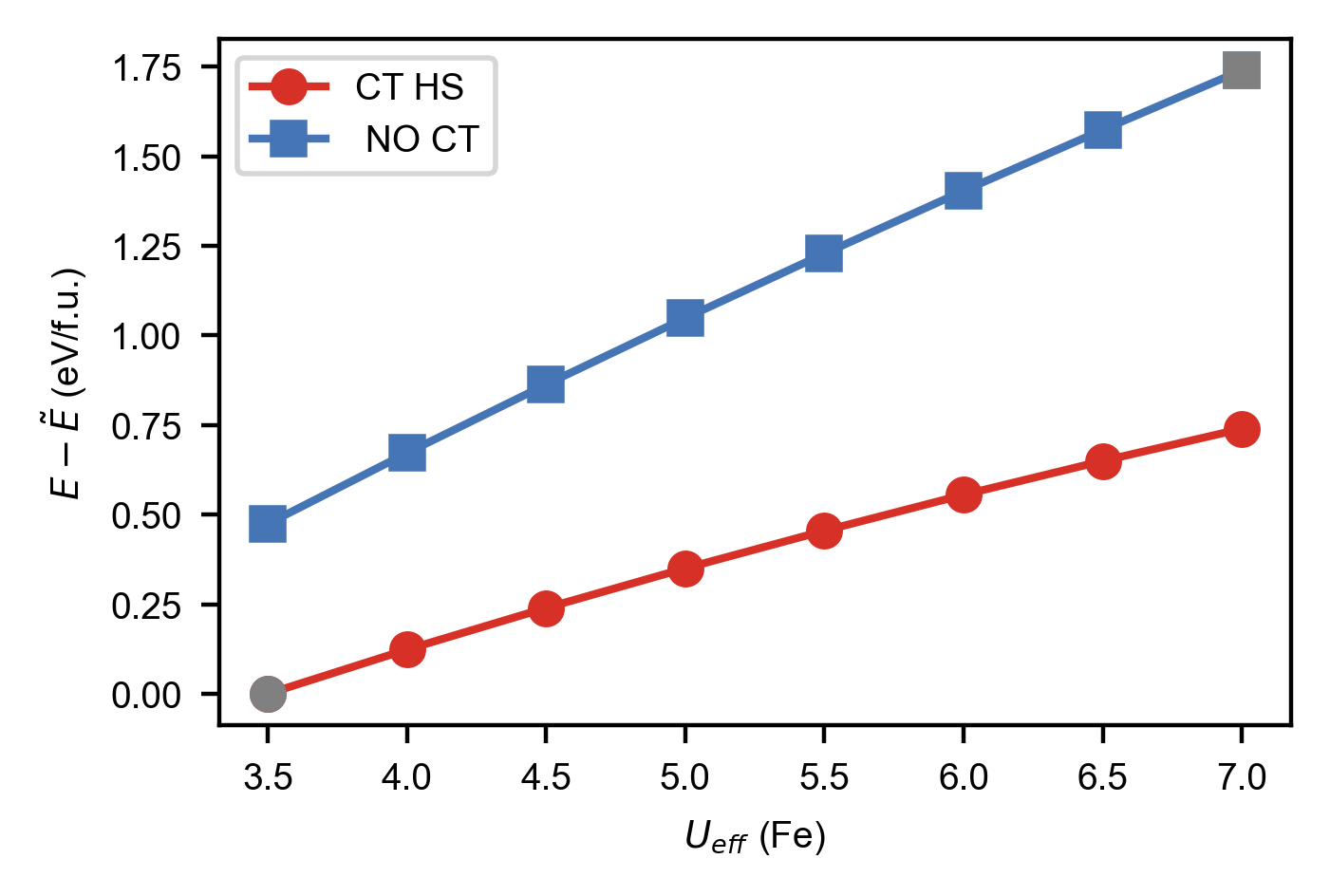}
\caption{\edits{Energy dependence of the two configurations of the \LTOLFO interface, which are charge-transfer high-spin (CT HS, with Fe$^{2+}_{HS}$ and Ti$^{4+}$) and no charge-transfer (NO CT, with Fe$^{3+}_{HS}$ and Ti$^{3+}$) configurations  as a function of different values of effective Coulomb repulsion for Fe, while the value on Ti has been fixed to $U_{eff}(Ti)=3 eV$. For plotting purposes, the energies are rescaled with respect to $\tilde E$ which is the lowest energy found in the shown data set.}}\label{fig:U_analysis}
\quad
\end{center}
\end{figure}

%%%%%%%%%%%%%%%%%%%%%%%%%%%%%%%%%%%%%%%%%%%%%%%%%
% FIG: multisubstrate CT HS Dos
%%%%%%%%%%%%%%%%%%%%%%%%%%%%%%%%%%%%%%%%%%%%%%%%%

\begin{figure}
\begin{center}
\hspace*{-0.65in}
\includegraphics[width=\textwidth,keepaspectratio]{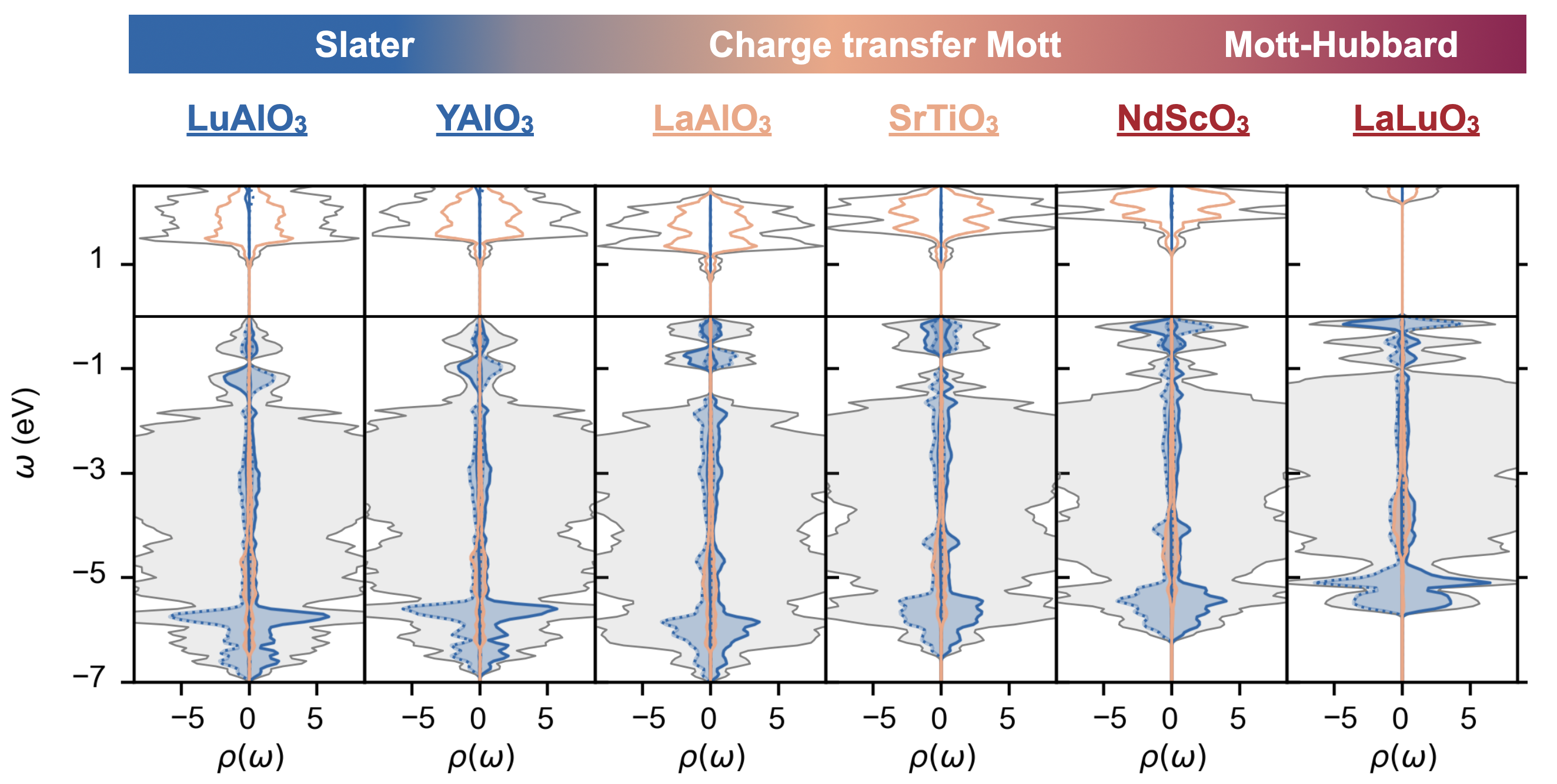}
\caption{Density of states of the charge transfer high spin (CT HS) configuration across all the substrates considered in our work with the Fermi level being shifted to the top of the valence band.The value of the band gap in function of the substrate is also shown. }\label{fig:Fig_S4}
\end{center}
\end{figure}

%%%%%%%%%%%%%%%%%%%%%%%%%%%%%%%%%%%%%%%%%%%%%%%%%
% FIG: AIRSS 
%%%%%%%%%%%%%%%%%%%%%%%%%%%%%%%%%%%%%%%%%%%%%%%%%

\begin{figure}
\begin{center}
\includegraphics{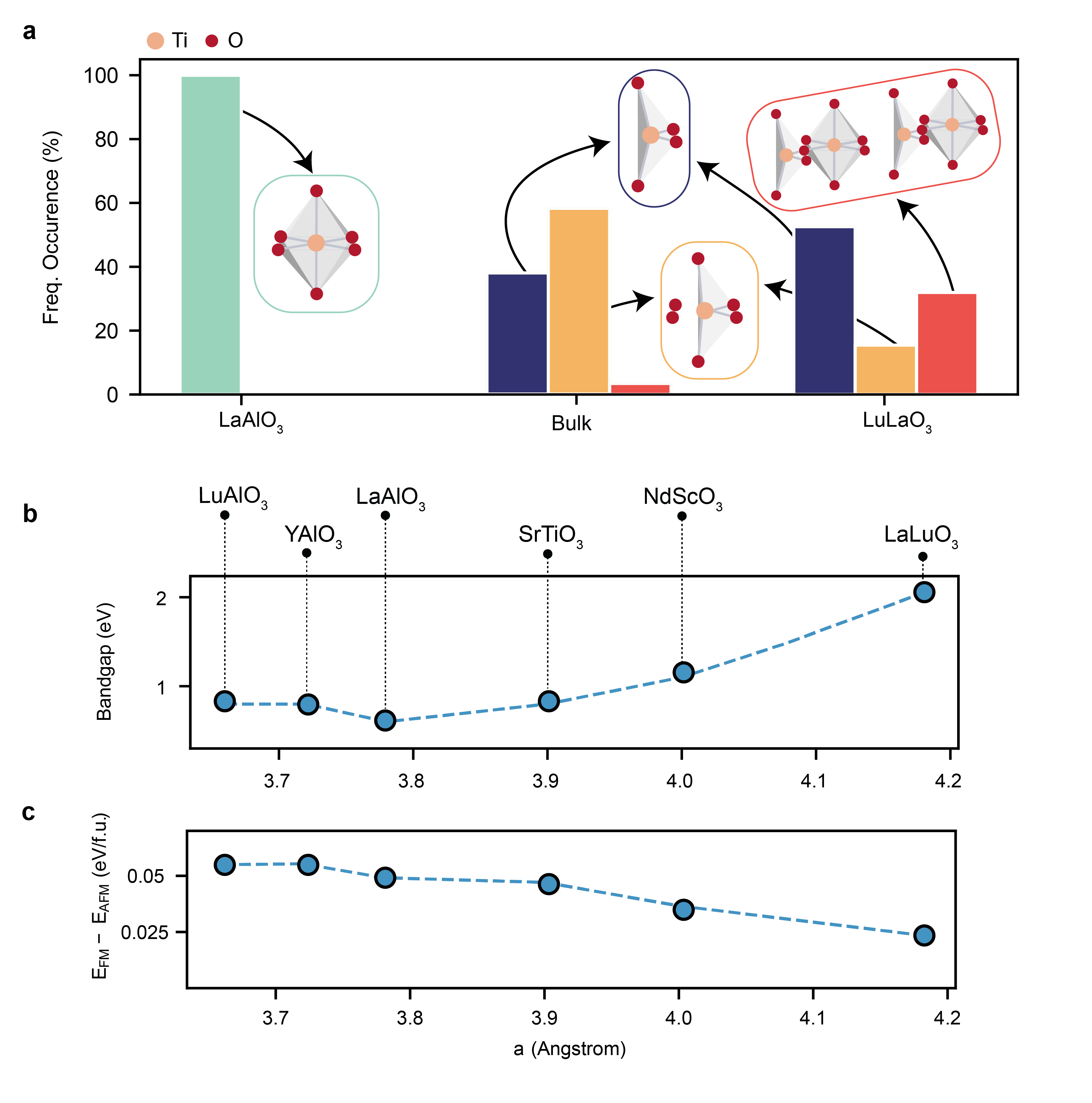}
\caption{(a) The local O-Ti-O coordinated phase space predicted by by spin assisted ab-initio random structure searches across the Slater to Mott transition. For each histogram the energetic ordering goes from left to right. (b) The substrate dependent behaviour of the bandgap across series of substrates examined. (c) The difference between the ferromagnetic and G-type I anitferromagnetic configurations across the studied substrates.}\label{fig:SM_Fig_2a}
\quad
\end{center}
\end{figure}

%%%%%%%%%%%%%%%%%%%%%%%%%%%%%%%%%%%%%%%%%%%%%%%%%
% FIG:Slater to Mott: Volumes and dos
%%%%%%%%%%%%%%%%%%%%%%%%%%%%%%%%%%%%%%%%%%%%%%%%%
\begin{figure}
\begin{center}
\includegraphics[width=\textwidth,height=0.65\textwidth,keepaspectratio]{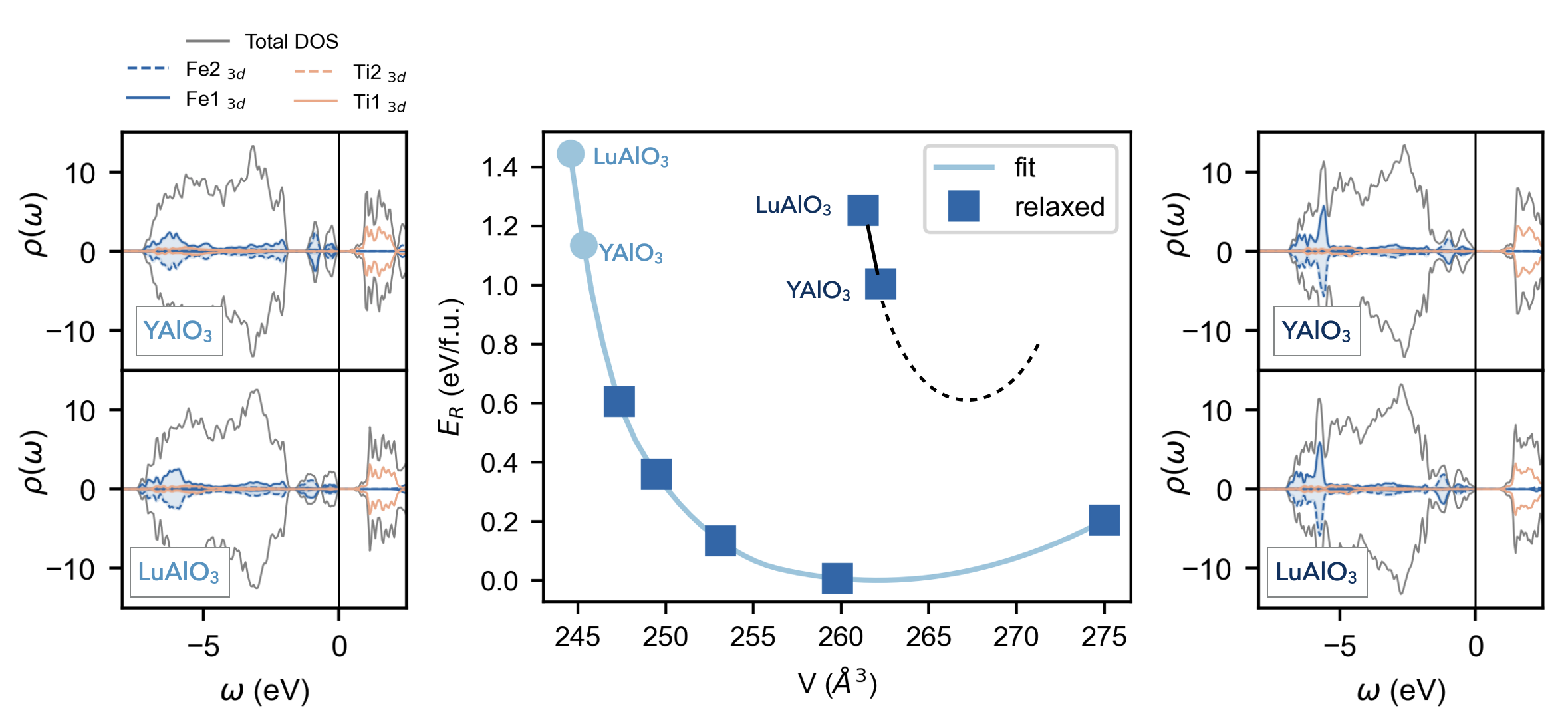}
\caption{Energy differences of the relaxed volumes (squares) of the \LTOLFO  heterostructure with charge transfer from Ti$^{4+}$ to Fe$^{2+}_{HS}$  configuration for different selected substrates (labels), with respect to the bulk (black star). For  \LuAO and \YAO substrates, fitted volumes are also shown (circles) with the associated density of states in the side panels. Indeed note that \LuAO and \YAO do not follow the same equation of state as the other substrates. Thus, two energy/volume profiles are added as guide to the eyes (CT HS (M) and CT HS (S)) corresponding to different character of the \eg orbitals, localised- Mott or itinerant-Slater, shown in the density of states (side panels).   The comparison of the energy of the two systems between the extrapolated volumes (red circles) and the relaxed ones (green squares), reveals that there is a further energy reduction when LAO and YAO expand their volume.}\label{fig:Fig_S6}
\quad
\end{center}
\end{figure}

%%%%%%%%%%%%%%%%%%%%%%%%%%%%%%%%%%%%%%%%%%%%%%%%%%%%%%%%%%%
% FIG: Slater to Mott: angles, polarization 
%        and crystal structure stability
%%%%%%%%%%%%%%%%%%%%%%%%%%%%%%%%%%%%%%%%%%%%%%%%%%%%%%%%%%%

\begin{figure}
\begin{center}
\includegraphics[width=\textwidth,keepaspectratio]{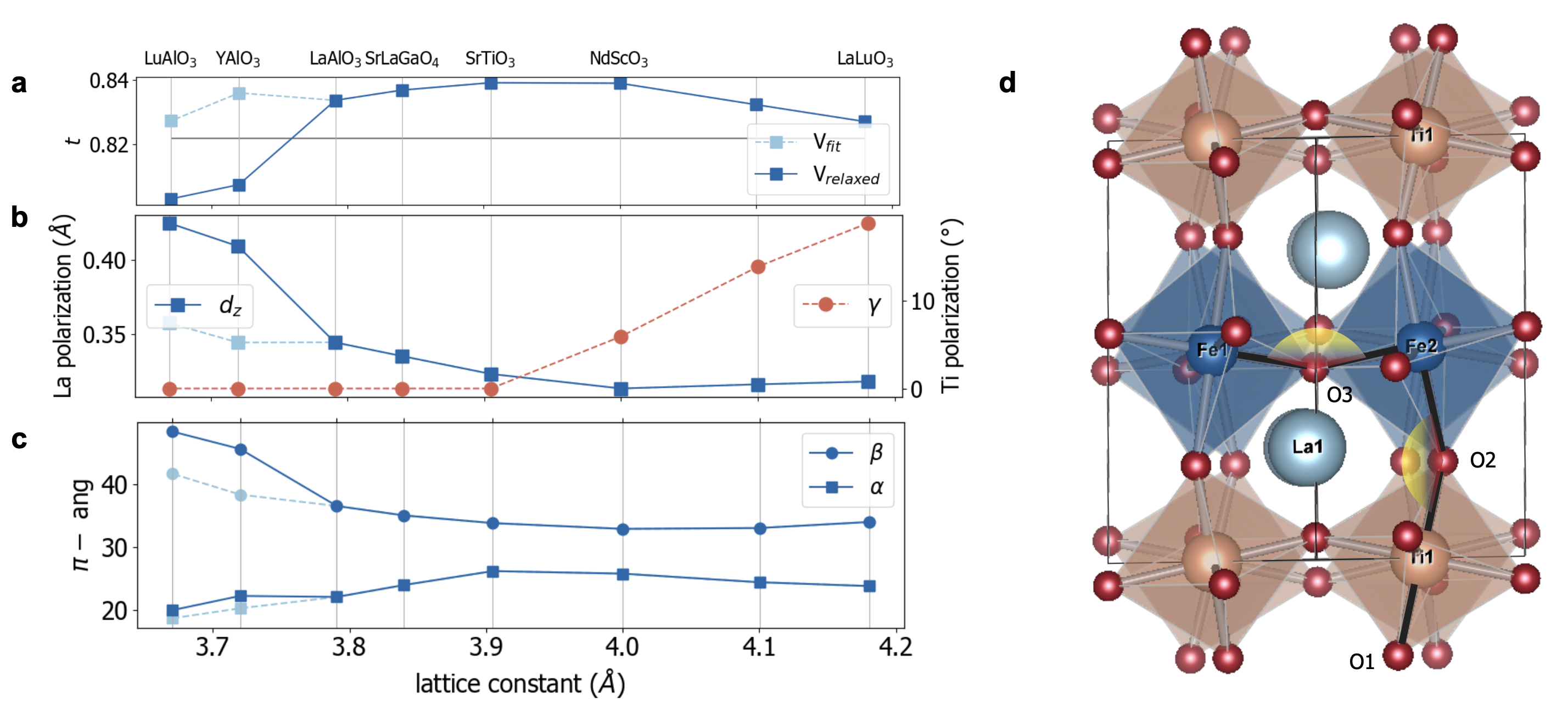}
\caption{Analysis of the structural degrees of freedom for  the \LTOLFO  heterostructure with charge transfer from Ti$^{4+}$ to Fe$^{2+}_{HS}$ (CT HS) configuration clamped on selected substrates (labels). In particular we show the behaviour of the Goldschmidt  tolerance factor (a), the La ($d_z$)and Ti ($\gamma$) polarization  (b) and the angles  (c) as defined in the structure shown in panel (d). In particular: $d_z$ is the offset along z direction  of the La from the La$_1$O$_2$ plane, $\gamma=\pi-\widehat{O_2Ti_1O_1}$, $\beta=\widehat{Fe_1O_3Fe_2}$ and $\alpha=\widehat{Fe_2O_2Ti_1}$  For  \LuAO and \YAO substrates, the quantities in analysis are computed for the fitted volumes (light blue) along the dashed line shown in Fig. (2b) in the main text. The gray solid line in panel b refers to the tolerance factor computed considering the ionic radii from \cite{Shannon}.}\label{fig:Fig_S7}
\quad
\end{center}
\end{figure}

%%%%%%%%%%%%%%%%%%%%%%%%%%%%%%%%%%%%%%%%%%%%%%%%
% FIG: DMFT occupations, G(iwn) and Sigma(iwn) %
%%%%%%%%%%%%%%%%%%%%%%%%%%%%%%%%%%%%%%%%%%%%%%%%

\begin{figure}
\begin{center}
\includegraphics[width=\textwidth,keepaspectratio]{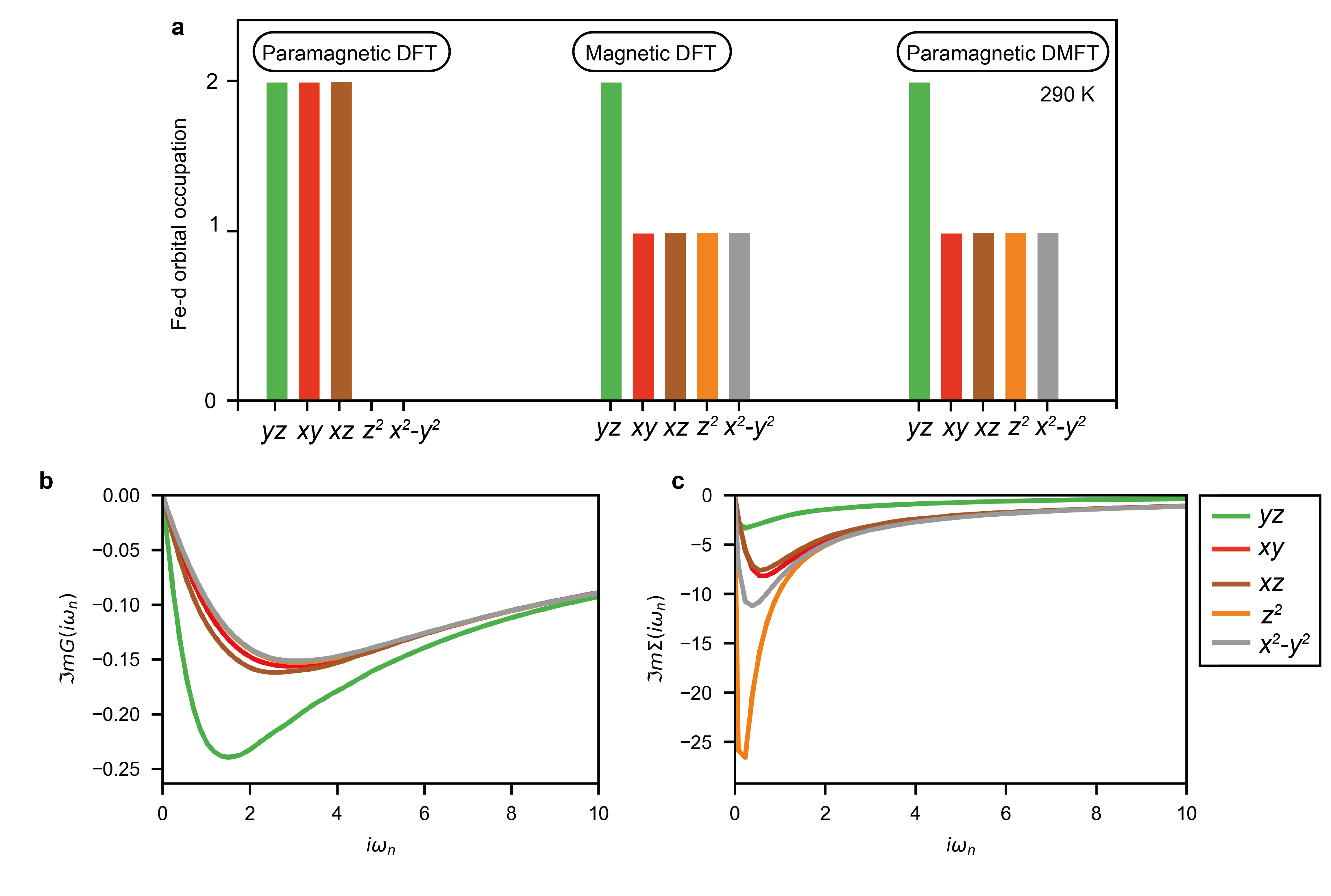}
\caption{\edits{(a) Histogram for the Fe-d orbitally resolved occupations on the LaLuO$_3$ substrate as predicted by \emph{(i)} paramagnetic DFT, \emph{(ii)} magnetic DFT and \emph{(iii)} paramagnetic DMFT at T=290 K (b)  Orbitally resolved imaginary part of the impurity green's function $\Im m G(i \omega_n)$ and (c) imaginary part of the self energy $\Im m \Sigma(i \omega_n)$ predicted at T=290 K.}}\label{fig:Fig_S8}
\quad
\end{center}
\end{figure}

%%%%%%%%%%%%%%%%%%%%%%%%%%%%%%%%%%%%
% FIG: Spectral function from DMFT %
%%%%%%%%%%%%%%%%%%%%%%%%%%%%%%%%%%%%

\begin{figure}
\begin{center}
\includegraphics[width=\textwidth,keepaspectratio]{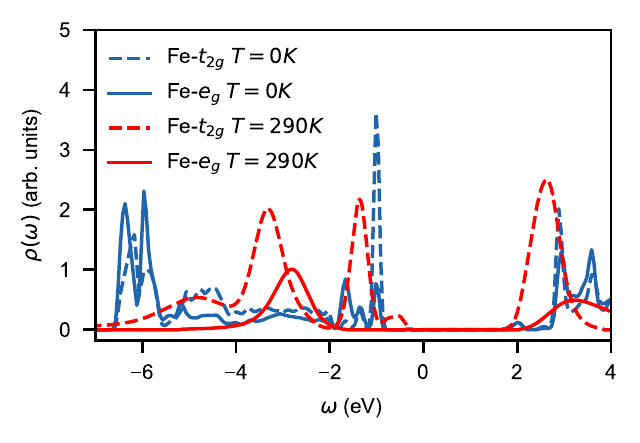}
\caption{\edits{Orbitally resolved Fe-d spectral function on the LaLuO$_3$ substrate decomposed along the octahedral crystal field axes with absolute spin components at T=0K (magnetic DFT) and T=290K (paramagnetic DFT+DMFT).}}\label{fig:Fig_S9}
\quad
\end{center}
\end{figure}

\begin{figure}
\begin{center}
\includegraphics[width=0.7\textwidth,keepaspectratio]{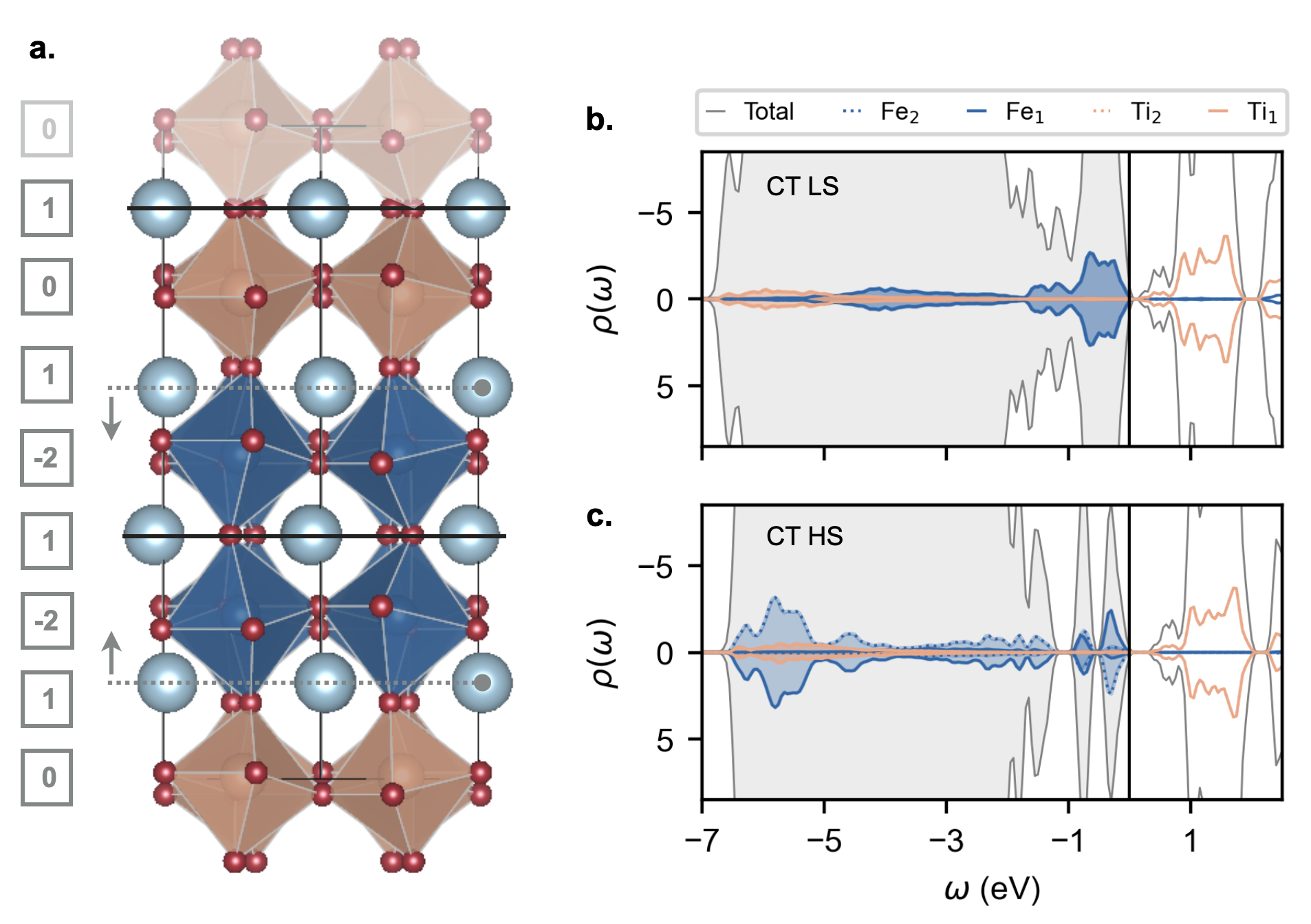}
\caption{\edits{\textbf{a.} (LaTiO3)$_2$/(LaFeO3)$_2$ heterostructure with charge transfer from Ti to Fe. The panel includes also the nominal charge in each layer and the alignment or displacement of the La ions from the oxygen plane toward the FeO$_2$ layers (as indicated by the arrows). Density of states obtained for the configurations with charge transfer and Fe$^{2+}$ being respectively in \textbf{b.} low spin (CT LS) and \textbf{c.} high spin (CT HS). The full Ti-d and Fe-d manifolds are shown and  are decomposed into their Fe$_{1,2}$ or Ti$_{1,2}$ spatial labels.} }\label{fig:22_supercell}
\quad
\end{center}
\end{figure}

%-----------------------------------------------------------------------------
%-----------------------------------------------------------------------------

\bibliographystyle{abbrv}
\bibliography{main.bib}